\documentclass{ws-ijmpa}
\usepackage[super,compress]{cite}
\usepackage{dcolumn}
\usepackage{bm}

\usepackage[utf8]{inputenc} 
\usepackage[T1]{fontenc}    
\usepackage{hyperref}       
\usepackage{url}            
\usepackage{booktabs}       
\usepackage{nicefrac}       
\usepackage{microtype}      
\usepackage{mathtools}
\usepackage{commath}
\usepackage[sc,osf]{mathpazo}

\usepackage{float}
\usepackage{varioref}
\usepackage[sans]{dsfont}
\usepackage{tikz}

\usepackage[makeroom]{cancel}
\usepackage{graphicx}

\newcommand{\Mathematica}{\textit{Mathematica\textsuperscript{\resizebox{!}{0.8ex}{\textregistered}}}}
\def\8{\infty}
\def\oh{\frac{1}{2}}
\def\ot{\frac{1}{3}}

\def\tt{\frac{2}{3}}
\def\ft{\frac{4}{3}}

\def\d{\partial}
\def\i{\imath\,}

\def\dal{\partial_{\alpha}}
\def\dbe{\partial_{\beta}}
\def\dga{\partial_{\gamma}}
\def\dla{\partial_{\lambda}}

\def\undertext#1{\vtop{\hbox{#1}\kern 1pt \hrule}}
\def\ra{\rightarrow}
\def\Ra{\Rightarrow}

\def\VEV#1{\left< #1\right>}

\def\pbyp#1#2{\frac{\partial#1}{\partial#2}}
\def\ff#1{\frac{\delta}{\delta#1}}
\def\fbyf#1#2{\frac{\delta#1}{\delta#2}}

\def\bea{\begin{eqnarray} & &}
\def\eea{\end{eqnarray}}

\def\EXP#1{\exp\left(#1\right)}
\def\INT#1#2{\int_{#1}^{#2}}
\def\PBR#1{\left[#1\right]}

\def\CL{Clebsch }

\def\PB{Poisson brackets }
\def\KO{Kolmogorov }
\def\NS{Navier-Stokes }

\def\BL {Beltrami}
\def \SYM {symplectomorphisms}
\def \GBF {$\mbox{GBF}$}
\def \PF {\mbox{pf }}
\def \DS {discontinuity surface}

\def\val{v_{\alpha}}
\def\vbe{v_{\beta}}
\def\vga{v_{\gamma}}

\def\ral{r_{\alpha}}
\newcommand{\Mod}[1]{\ (\mathrm{mod}\ #1)}
\def\rbe{r_{\beta}}

\def\oal{\omega_{\alpha}}
\def\obe{\omega_{\beta}}
\def\oga{\omega_{\gamma}}

\def\XXint#1#2#3{{\setbox0=\hbox{$#1{#2#3}{\int}$}
     \vcenter{\hbox{$#2#3$}}\kern-.5\wd0}}

\newcommand{\Z}{\mathbb{Z}}

\DeclareMathOperator*{\argmin}{arg\,min}

\DeclareMathOperator{\arcsinh}{arcsinh}
\DeclareMathOperator{\arctanh}{arctanh}

\DeclareMathOperator{\erf}{erf}
\DeclareMathOperator{\Li}{Li}

\usepackage{comment}
\begin{document}


\title{Clebsch Confinement and Instantons in Turbulence}%
\author{Alexander Migdal}
\address{Department of Physics, New York University \\
  726 Broadway, New York NY 10003}

 




\date{\today}
\maketitle

\begin{abstract}
    The Turbulence in incompressible fluid is represented as a Field Theory in 3 dimensions. There is no time involved, so this is intended to describe stationary limit of the Hopf functional. 
    The basic fields are Clebsch variables defined modulo gauge transformations (symplectomorphisms).
    Explicit formulas for gauge invariant Clebsch measure in space of Generalized Beltrami Flow compatible with steady energy flow are presented.
    We introduce a concept of Clebsch confinement related to unbroken gauge invariance and study Clebsch instantons: singular vorticity sheets with nontrivial helicity. This is realization of the "Instantons and intermittency" program we started back in the 90ties\cite{FKLM}.
    These singular solutions are involved in enhancing infinitesimal random forces at remote boundary leading to critical phenomena. In the Euler equation vorticity is concentrated along the random self-avoiding surface, with tangent components proportional to the delta function of normal distance.
    Viscosity in Navier-Stokes equation smears this delta function to the Gaussian with width $h \propto \nu^{\nicefrac{3}{5}}$ at $\nu \ra 0$ with fixed energy flow.
    These instantons dominate the enstrophy in dissipation as well as the PDF for  velocity circulation $\Gamma_C$ around fixed loop $C$ in space.
    At large loops, the resulting symmetric exponential distribution perfectly fits the numerical simulations\cite{IBS20} including pre-exponential factor $1/\sqrt{|\Gamma|}$.
    At small loops, we advocate relation of resulting random self-avoiding surface theory with multi-fractal scaling laws observed in numerical simulations. These laws are explained as a result of fluctuating internal metric (Liouville field). The curve of anomalous dimensions $\zeta(n)$ can be fitted at small $n$ to the parabola, coming from the Liouville theory with two parameters $\alpha, Q$. At large $n$ the ratios of the subsequent moments  in our theory grow linearly with the size of the loop, which corresponds to finite value of $\zeta(\infty)$ in agreement with DNS.
    
\end{abstract}


\section{\label{sec:level1} Introduction: Waves vs Instantons}
Allegedly Richard Feynman said ``Turbulence is the most important unsolved problem of classical physics.'' He may have indeed said that in 1970 but it was not published by him, so we rely on second-hand quotes\cite{EF11}.

The only published quote I found was in ``Feynman's Lectures in Physics'' \cite{Feynman} first published in 1963, and it is much deeper:

``Finally, there is a physical problem that is common to many fields, that is very old, and that has not been solved. 

It is not the problem of finding new fundamental particles, but something left over from a long time ago—over a hundred years. Nobody in physics has really been able to analyze it mathematically satisfactorily in spite of its importance to the sister sciences. 

It is the analysis of \textit{circulating or turbulent fluids}. 

If we watch the evolution of a star, there comes a point where we can deduce that it is going to start convection, and thereafter we can no longer deduce what should happen. A few million years later the star explodes, but we cannot figure out the reason. 

We cannot analyze the weather. We do not know the patterns of motions that there should be inside the earth. The simplest form of the problem is to take a pipe that is very long and push water through it at high speed. We ask: to push a given amount of water through that pipe, how much pressure is needed?

No one can analyze it from first principles and the properties of water. If the water flows very slowly, or if we use a thick goo like honey, then we can do it nicely. You will find that in your textbook. What we really cannot do is deal with actual, wet water running through a pipe. 

That is the central problem which we ought to solve some day, and we have not.''\footnote{I am glad that he mentioned sister sciences, as I am going to use here the sister Quantum Field Theory with its functional integrals, initiated by Feynman. I am also glad he mentioned the circulating fluid, as velocity circulation plays the major role in my theory.}

Another half century passed since he wrote this, and we still have not solved it.

By solution of this problem he meant mathematical description of statistics of the turbulent flow from the first principles, which is Navier-Stokes equation
\begin{eqnarray}\label{NSv}
   &&\d_t \val = \nu \dbe^2\val - \vbe \dbe \val - \dal p ;\\
   && \d^2 p + \dal \vbe \dbe \val =0;
\end{eqnarray}
The second equation here reflects the fact that the fluid is incompressible, so that the pressure instantly adjusts to velocity evolution by providing conservation of incompressibility condition
\begin{equation}
    \dal \val =0
\end{equation}

In this work we are only considering the real world with three dimensions. Turbulence in other dimensions may be quite different, in particular odd and even dimensions have different topological invariants. But Feynman had three dimensions in mind and so shall we.

This equation is deceptively simple which makes the problem so appealing. The problem is that this equation does not have a stable smooth solution given large enough energy flow into the fluid. 

This unstable solution is not unique.
It can be described as statistical distribution of velocity field which distribution is believed to be universal in the infinite volume. It is observed in myriad natural phenomena starting with the water flowing from your faucet and ending with mega-parsec turbulence in the Universe.

This statistical distribution represents a steady state in a sense that all the energy pumped into the flow by external forces from the boundary is dissipated inside the fluid. Nobody have provided the microscopic definition of this distribution, unlike the Gibbs distribution in statistical physics.

The problem looks analogous to critical phenomena in statistical physics, but there are important distinctions. The critical phenomena, as we know for the last 40 years, are essentially local -- there is conformal invariance corresponding to local scale transformations of fluctuating fields.

The conformal invariance uniquely fixes the scaling dimensions $\Delta = d-1$ of conserved vector fields in $d$ dimensions like velocity here. This is very far from observed scaling laws with $\Delta \approx -\ot$ in turbulence, so the velocity cannot be a conformal field.

Also, the vorticity
\begin{eqnarray}
   &&\oal = e_{\alpha\beta\gamma} \dbe \vga;\\\label{OV}
   && \dal \oal =0
\end{eqnarray}
which is also conserved, being the derivative of velocity, has dimension $\Delta +1$, which is another contradiction. None of these two conserved fields can be a conformal field with any dimension.

The local scaling symmetry is broken by the pressure, which is a non-local functional of velocity field, obtained by solving the Poisson equation in \eqref{NSv}. This equation is not conformal invariant. 

There are other puzzling features of the \NS{} equation. 
In the limit of vanishing viscosity $\nu = 0$ this equation becomes Euler equation, which is time-reversible. Still, the dissipation does not go away at arbitrary small viscosity. This is viscosity anomaly we discuss later in great detail.

Mathematically, of course this means that this limit $\nu \ra 0$ cannot be uniform in space. The viscous term $ \nu \dbe^2\val$ in \eqref{NSv} has more derivatives that nonlinear Euler term $\vbe \dbe \val $. These terms could balance in the limit $\nu \ra 0$ if the velocity field is not smooth, at least in some regions in space.

In conventional approach to Turbulence, where velocity field is the basic fluctuating variable, there are singular correlation functions such that the singularities at coinciding points in the chain of steady state equations for these correlation functions compensate for small value of viscosity\cite{SY05}.

Effective UV cutoff length (viscous scale) in these scaling models goes to zero with viscosity and negative powers of this viscous scale compensate small viscosity, so that the dissipation persists.

In particular, there is a famous Kolmogorov law (with $ d = 3$ being the space dimension and $V$ being the total volume)
\begin{eqnarray}\label{KOLM}
    &&\VEV{\val(0) \vbe(0) \vga(r)} =
\frac{\mathcal E }{V(d-1)(d+2)}
	\left(
	 \delta_{\alpha \gamma} r_{\beta} +
	 \delta_{\beta \gamma} r_{\alpha} -
	 \frac{2}{d}\delta_{\alpha \beta} r_{\gamma}
	\right);\\
	&& \mathcal E  = \VEV{\int_r \val \vbe \dbe \val}
\end{eqnarray}
which explicitly violates the time reversal symmetry.

There is also the exact relation for the energy dissipation (equal to the energy flow) in homogeneous turbulence
\begin{eqnarray}
   \mathcal E  = \nu \VEV{ \int_r \oal^2}
\end{eqnarray}

Here, in the limit $\nu \ra 0$ the so called enstrophy $\VEV{ \int_r \oal^2}$ must grow to compensate the factor of $\nu$. Usually this is explained \cite{Y07} by splitting points in $ \oal^2 \Ra \oal(0) \oal(r)$ and cutting off the singular power law at viscous scale.

In any case we see that the relevant velocity fields are not smooth, creating some UV divergences leading to viscosity anomaly.

Statistics of velocity differences or vorticity fields as measured in numerical simulations as well as real experiments is far from Gaussian. Local statistics of velocity field is numerically close to Gaussian, but this is beside the point. The effective Hamiltonian for velocity field is non-local and non-Gaussian.

By all standards this is a strong coupling phase of whatever field theory describes the velocity fluctuations.

The so called multi-fractal scaling laws suggested by Parisi and Frisch in the late 80-ties\cite{FP85} as fitted to the measured moments of velocity field differences\footnote{We make distinction between longitudinal and transverse components of velocity differences. The correct combination from our point of view is $\delta_i v_j - \delta_j v_i$ which represents the circulation over a square. The potential components, which drop here, are not local and so are much more complex in our theory.}
\begin{equation}\label{Zeta}
    \VEV{(v_\perp(\vec r) - v_\perp(0))^n} \propto |\vec r|^{\zeta(n)}
\end{equation}
showed nonlinear anomalous dimension $\zeta(n)$ \cite{DS98, Y07}, growing with $n$ and reaching a plateau (see\cite{ISY20} for recent large scale DNS).  

One could also interpret these results for $\zeta(n)$ as a sequence of transitions at various Reynolds numbers\cite{Y14}, except for exact Kolmogorov value $\zeta(3) =1$ which follows from \eqref{KOLM}.

Positive values of these $\zeta(n)$ correspond to velocity correlations growing with distance, in contrast with decreasing correlations of CFT in critical phenomena.

These growing correlations reflect some coherent vorticity structures -- vortex cells\cite{TSVS} which were observed in numerical simulations (see\cite{Hal20} for the recent review) as well as real experiments\cite{IM19}. 

Such coherent structures cannot be described as collection of waves in the same way as the QCD string cannot be described as collection of gluons. 

The Fourier transform hides these structures by imposed periodicity. Recovering these structures from the Fourier analysis is as hopeless as recognising the shape of mysterious smile of Mona Lisa from the color spectrum of the paint.

The perturbation theory (or, in general, WKB expansion around some smooth potential flow), which is our only analytical tool in field theory, fails here.  It leads to divergent expansion in inverse powers of viscosity.

This strong coupling problem tortured us for 80 years since the Kolmogorov-Obukhov discoveries in 1941. Not a single exact theoretical formula was discovered from first principles in 3D Turbulence after that, though there is abundance of numerical results and some phenomenological models\cite{Y07,Y14}, working quite well in engineering applications, like Feynman's example of computation of the pressure needed to push water through the pipe.

In particular, in \cite{SY05, Y07} the balance of viscosity and nonlinear terms in \NS{} equation was investigated for a chain of the Hopf equations for the velocity correlation functions, using some model for the pressure as function of velocity to close this set of equations. 

Within this model, the viscosity anomaly (persistence of the dissipation in the limit of zero viscosity) was explained  in terms of the singular correlation functions.

The phenomenological models are useful in practical applications, but still we need to understand the microscopic mechanisms of turbulence, reveal the hidden structures and hidden laws of their statistics, if not for the engineering needs then just out of eternal scientific curiosity.

There were, meanwhile, some important observations made on purely theoretical level as well. The topologically conserved helicity integral
\begin{eqnarray}
   H = \int_r \oal \val
\end{eqnarray}
indicates some nontrivial knotting of vortex lines \cite{MO69}. The relevance of this invariant to turbulence was advocated in Kraichnann's helical turbulence \cite{KR73}, though this has not led to a quantitative theory then. 

The dynamics of knotted vortexes in turbulent fluids was studied in real experiments in \cite{KI13} in a turbulent water. These authors observed the conservation of helicity and reconnection of vortex lines due to viscous effects. 

The relation of helicity to the \CL{} topology was pointed by \cite{KM80} and \cite{L81} and then used in my early attempts \cite{TSVS} to connect the turbulence to the random surface theory.

\textbf{The main idea of my approach, as started in the 90ties and recently advanced in 2019-20 is that there is a dual, geometric view on statistics of Turbulence. Instead of strongly interacting nonlinear waves we are looking for weakly interacting singular vorticity sheets -- instantons. The viscosity anomaly is then explained by surface singularities, in the same way as quark confinement in QCD was explained by gauge field collapsing to the minimal surface inside Wilson loop. This dual view complements rather than contradicts the old view, and it dramatically simplifies in the limit of large circulations.}

The velocity discontinuities (shocks) are known to exist in Burgers turbulence \cite{BK07}, which is a one-dimensional toy model for a turbulence. The exponential tails of velocity difference PDF which was explained in \cite{GM96} as an instanton dominance, was also explained in terms of these discontinuities.

In a retrospect, these shocks in Burgers turbulence should have inspired the search of similar shocks in three dimensional Euler equations, but this is not what have driven me. I simply forgot about these one-dimensional velocity discontinuities and arrived to my \DS{} from a different angle, related to helicity. Now I clearly see that analogy between shocks and instantons.

Our singular surfaces arise as discontinuities of velocity field in physical space in the limit of vanishing viscosity. The normal velocity linearly vanishes  at this singular surface, but the tangent velocity has a finite jump. The normal component of vorticity is finite at the surface, but tangent components are proportional to the delta function of the normal coordinate $z$.

This singular vorticity in three dimensional turbulence is smeared by viscosity, so that at small viscosity we have so called Zeldovich pancake\cite{SZ89}: the thin layer of large vorticity, corresponding to smeared discontinuity of velocity in the region $|z| \sim h(\nu)$. 

The thickness $h(\nu)$ goes to zero as some power of $\nu$. These peaks of vorticity lead to the viscosity anomaly in dissipation $ \int_r \oal^2 \sim \nu^{-1}$.

The regions of high vorticity were observed in numerical simulations staring with She, Jackson and Orszag\cite{She1990}. Recently, these regions were studied in large scale numerical simulations in \cite{Buaria_2019}. These regions form all kind of shapes, some are tubes, other are like sheets. 

Recently \cite{ISY20}, some additional numerical evidence for vorticity sheets was presented and used to explain the saturation of transverse scaling exponent $\zeta(\infty)\approx 2$.

All of these shapes are candidates for our singular vorticity surfaces, some closed, other bounded by fixed loops in space in case of PDF for velocity circulation.

There are topological reasons  (winding  of \CL{} field on unit sphere $S_2$) for these discontinuities and singularities which we discuss in detail later.

If we simply assume such a smeared discontinuity it is not difficult to see from \eqref{NSv} that the vorticity shape must be Gaussian, and velocity discontinuity is smeared to the error function.

Let us give a hint how these smeared singularities arise before systematically studying them in the rest of the paper.

We are considering vicinity of the \DS{} with local $x,y$ tangent plane and local normal direction $z$. Let us study the most singular terms, involving derivatives in normal direction in the \NS{} equation.

We assume the Euler discontinuity in tangent components $v_i = (v_x, v_y)$ of velocity. As for the normal component $v_z$ it must go to zero at $z\ra 0$, and we assume that it goes to zero linearly with $z$, as there are no singularity in normal velocity in these Euler solutions (as we shall study in detail later).

The tangent velocity, including its discontinuity, in general varies along the surface, as we shall study in detail in this paper. Only the \CL{} field $\phi_2$ has constant discontinuity $ 2 \pi n$ related to its winding number.

The viscous term at small $z$ in this equation would go as $\nu \d_z^2 v_i$, with $v_i$ being the tangent components. The Euler term would go as $v_z \d_z v_i$.  As $v_z \ra 0$ we have $v'_z z \d_z v_i$. 
Matching these two terms\footnote{Sreenivassan and Yakhot \cite{SY05}  were also matching contributions from viscous and nonlinear terms in  \NS{} equation to the moments of velocity differences. They considered homogeneous turbulence with singular correlations so they were led to different matching models. We are using the dual view of the singular surfaces, so we match these two terms in the vicinity of the \DS{} without any assumptions about correlation functions or closure of moments. Our approach is much simpler and it does not need any assumptions, except for existence of the \DS{}. The duality of Turbulence we advocate in this paper means that \textbf{both} views are valid, they complement rather than contradict each other.} leads to the equation (with $v_i$ depending on all $x,y,z$)
\begin{equation}
   \nu \d_z^2 v_i = v'_z z \d_z v_i
\end{equation}
which has singular solution we need
\begin{equation}
   v_i \propto \erf\left(\frac{z}{h\sqrt 2}\right);
\end{equation}

The corresponding vorticity behaves as a Gaussian with width $h$
\begin{equation}
      \omega_i \propto \frac{1}{h} \EXP{- \frac{z^2}{2 h^2}};
\end{equation}
There are smooth functions of the surface point $x,y$ in front of these $z-$ dependent factors in velocity and vorticity.

The normal derivative of normal velocity is related to the thickness $h$
\begin{equation}
   v'_z = \frac{\nu}{h^2}
\end{equation}
Note that this means that this normal derivative of normal velocity is constant along the discontinuity surface, unlike the tangent components of velocity.

By naive estimate this would mean that $h \sim \sqrt{\nu}$ but more careful analysis shows that in order to have finite energy flow in viscous anomaly $\nu \int_r \oal^2$ the width $h$ should go to zero slightly faster, as $\nu^{\frac{3}{5}}$. In the turbulent limit of $\nu \ra 0$ at fixed energy dissipation we recover delta function singularity in tangent vorticity and the discontinuity in tangent velocity.

Note that unless $v_z =0$ at $z=0$ there is no solution bounded on both sides of the surface-- it exponentially grows on one side regardless of the sign of $v_z$. Further investigation using Clebsch variables reveals that this discontinuity is proportional to a certain winding number $n$ related to the helicity.

In general, the surface of discontinuity is arbitrary, so that winding numbers, positions, sizes and shapes of these surfaces represent the degrees of freedom in our statistical distribution. 
In case of PDF for velocity circulation around the large loop fixed in space, however, all these degrees of freedom freeze so we are left with the minimal surface bounded by the loop and the lowest winding number $n = \pm 1$.

This is the main result of our research. The rest are technical details, topological arguments and some computations of observables based on these singular flows.

The predictions are quite specific and verifiable. In particular, the PDF for velocity circulation $\Gamma$ around large loop goes as sum of exponential terms $|\Gamma|^{-\oh}e^{- n b |\Gamma|}$. 

The leading term with $n=1$ fits numerical simulations\cite{S19,IBS20,M20c} with high precision over six decades of exponential decay, including time reversal symmetry symmetry $\Gamma \Ra - \Gamma$ and pre-exponential factor. The sub-leading corrections with $ n >1$ have not yet been observed.

\section{Hopf Equation for vorticity}

Let us introduce and study the Hopf functional.
 Navier-Stokes equation can be rewritten as equation for vorticity
\begin{subequations}
\begin{eqnarray}\label{NSE}
    &&\d_t \omega_\alpha = G_\alpha[\omega];\\
    && G_\alpha[\omega] = \nu \d^2\oal + \obe \dbe \val - \vbe \dbe \oal;
\end{eqnarray}
\end{subequations}
As for velocity, it is a given  by a Biot-Savart integral
\begin{equation}\label{BS}
    \val(r)= -e_{\alpha\beta\gamma} \dbe \int d^3 r' \frac{\oga(r')}{4\pi | r - r'|}
\end{equation}
which is a linear functional of the instant value of vorticity.

In conventional approach to the Turbulence there are Gaussian random forces concentrated on the large wavelengths.  These forces are usually added to the right side of \NS equation for velocity field. The Gaussian functional integral for these forces after inserting \NS equation as a condition with Lagrange multiplier leads to the Wylde functional integral, which involves time dependent fields. 

This functional integral in addition to providing the perturbation expansion in inverse powers of viscosity allowed some non-perturbative solutions\cite{FKLM} which were called instantons in analogy with the same non-perturbative solutions in gauge field theories. Explicit solutions were found for passive scalar\cite{FKLM} and Burgers equation\cite{GM96}.

Unfortunately, the attempt to find relevant instantons for the full \NS{} equation for velocity field failed. The only solution found in\cite{FKLM} described PDF falling faster than exponential decay $\EXP{-a \left(\delta v\right)^3}$ in strong disagreement with experiments. This solution was smooth and had no helicity.

We think that the root cause was the wrong variable choice. The velocity field and its fluctuations are influenced by the external forces, and its potential component has nontrivial dynamics. However, hidden deep inside this dynamics there is much simpler dynamics for the \CL{} variables. These variables can have nontrivial topology which was necessary for existence and stability of the instantons in the 2D sigma model and 4D gauge theory. 

As it was observed in my recent work\cite{M20a,M20b} one can provide energy flow to the bulk of the fluid from its boundary by purely potential forces $ f_\alpha =  -\dal \tilde p$. In\cite{IM19} similar conditions were achieved in real water: the forcing came from the corners of a large glass cube and the turbulence was confined to a blob in the center of that cube, far away from the forcing. 

Such purely potential random forces will drop from the right side of equation for vorticity. The restriction of the fixed energy flow, coming from the velocity equation, becomes a global constraint on our vorticity dynamics.

There is only one way these forces can influence vorticity: through the boundary conditions at infinity. Velocity in the bulk of the turbulent flow, where vorticity is present, depends of these random forces acting at infinity as a boundary condition for the pressure. This velocity moves vortex structures around and this is how the random forces influence vorticity dynamics.

The generating functional for single time vorticity distribution
\begin{eqnarray}
    &&H[\vec \lambda,t] = \VEV{\EXP{\i \int_r  \lambda_\alpha \oal}};\\
    && \lambda_\alpha = \lambda_\alpha(\vec r);\\
    &&\oal = \oal(\vec r, t)
\end{eqnarray}
is known to satisfy the Hopf equation\cite{Hopf19}:
\begin{eqnarray}\label{Hopf}
    \d_t H = \i \int_r\lambda_\alpha G_\alpha\left[-\i \ff{\lambda}\right] H
\end{eqnarray}
with averaging over randomized initial conditions being implied.

The vorticity PDF is given by functional Fourier transform (with $\oal=\oal(\vec r)$ being time independent variable)
\begin{eqnarray}
   P[\oal,t] = \int D \lambda \EXP{-\i \int_r \lambda_\alpha \oal}H[\vec \lambda,t]
\end{eqnarray}

As it is, the Hopf equation describes decaying turbulence, because of the dissipation in the \NS{} operator $G_\alpha[\omega]$. However, if we switch from initial conditions to the boundary conditions at infinity, providing constant energy flow, this equation could in principle have a steady solution, in other words a fixed point. 

The averaging $\VEV{}$ in this case becomes an averaging over these boundary conditions with mean energy flow staying finite and positive.

This averaging over boundary conditions means the following. Pick a realization of random force on a large bounding sphere taken from Gaussian distribution
with zero mean and finite variance. Solve the Hopf equation with this boundary force (time independent, but randomly chosen from a distribution).

Solve it again many times for different realizations of random forces. The Hopf equation being \textbf{linear}, the mean value of these Hopf functionals would be equivalent to integrating it over forces with some distribution. 

This method offers an alternative to traditional study of Turbulence by time averaging of stochastic differential equation (\NS{} with time-dependent Gaussian random forces). Time average of a  generating functional over Gaussian random variables with correlation $\propto K(\vec r - \vec r')\delta(t-t')$ is equivalent to averaging over ensemble of Gaussian forces with correlation $\propto K(\vec r - \vec r')$. Without correlations at $t \neq t'$ these forces at different times in stochastic differential equation are just independent samples from the same static Gaussian distribution.

The actual time dynamics may be needed to study kinetic phenomena, but not the single time statistics, which is given by steady state solution of the Hopf equation. This is what worked so well for centuries in ordinary statistical mechanics after the Gibbs fixed point was discovered. Dropping one of four variables in the equation is a big simplification of mathematical problem, not to mention a discovery of a new law of Physics. 

\section{Fixed Point}
Let us consider a manifold $\mathcal{G}$ of locally steady solutions (generalized Beltrami flow, \GBF )
\begin{eqnarray}
    \mathcal{G}: G_\alpha[\omega^\star,\vec r] =0
\end{eqnarray}
Then an integral
\begin{eqnarray}
    &&H \propto \int_\mathcal{G} d \mu(\omega^\star) \EXP{\i \int_r  \lambda_\alpha \omega^\star_\alpha};\\
    &&P \propto \int_\mathcal{G} d \mu(\omega^\star) \delta\left(\omega - \omega^\star\right)
\end{eqnarray}
with some invariant measure $d \mu(\omega^\star)$ on $\mathcal{G}$ would be a fixed point of  the Hopf equation as one can check by direct substitution into \eqref{Hopf}.

The random initial and boundary conditions are hidden in the distribution $d \mu(\omega^\star)$ in this formula. As we shall discuss in detail later, in addition to the local variables parametrizing vorticity $\omega^\star$ there are some global parameters which are also distributed with some weight. 

That includes uniform random forces, represented by just three global Gaussian variables $\vec f$. In addition, there is a global scale variable $Z$ which is involved in energy flow distribution.

As for the source $\lambda_\alpha(\vec r)$ we restrict ourselves to the function concentrated on a surface $S_C$ bounded by some loop $C$ in space
\begin{eqnarray}
    &&\lambda_\alpha(\vec r) = \gamma \int_{S_C} d \sigma_\alpha(\vec r') \delta^3(\vec r - \vec r') ;\\
    && \int d^3 r \lambda_\alpha(\vec r) \oal(\vec r,t) =  \gamma \Gamma_C;\\
    && \Gamma_C =  \int_{S_C} d \sigma_\alpha(\vec r') \oal(\vec r') = \oint_C \val d \ral
\end{eqnarray}

This way, our Hopf functional becomes the generating functional for the distribution of velocity circulation $\Gamma_C$. The loop equations\cite{M93,M19a,M19b,M19c,M19d} represent a specific case of the Hopf equation for this generating functional as a functional of the shape of the loop $C$. We do not need these equations in this work, though they were instrumental in derivation of Area law which we independently confirm.

This is the program we are implementing in our recent papers: we construct invariant measure on this manifold of \GBF{} and we  study the tails of PDF which as we argue are dominated by singular flows in Euler limit (smeared at viscous scales in full \NS ).

The viscous term $\nu \d^2\oal$ in \NS{} equations does not go away in the turbulent limit $\nu \ra 0$, apparently because of some singular configurations with infinite second derivatives of vorticity in the Euler equation. Would it go away, the turbulence would be time-reversible, contrary to all observations.\footnote{Strictly speaking, as we shall see below, the viscous effects \textbf{do} go away in extreme turbulent limit, as the peak in vorticity approaches the delta function, the thickness of Zeldovich pancake goes to zero and circulation scale goes to infinity.}

Numerous DNS support this viscosity anomaly phenomenon (\cite{S19} and references therein). My attention was attracted recently by an unpublished work\cite{S17} where various terms in the vorticity equation as well as correlations between them were investigated.

This DNS as well as all the rest, was dealing with  steady state of the forced \NS{} equation, where mean value $\VEV{\d_t \omega} $ vanished.  They observed that in this steady state, the balance of the terms indicated that the flow was far from the Euler steady state where $\obe \dbe \val = \vbe \dbe \oal$.

This could only mean that the viscous term remained significant in the turbulent limit. At the same time the magnitude of random forces presumably goes to zero in this limit, as  the nonlinearities of the \NS dynamics magnify the random fluctuations leading to finite energy flow.

This fixed point of the Hopf evolution is a candidate for the Turbulence statistics, but is it the right one? We can find out by investigating this distribution on theoretical level and comparing it with numerical simulations of the \NS{} equation. 

In the same way as with critical phenomena in ordinary statistical physics, we expect Turbulence to be universal \footnote{A good lesson of such universality was the description of 2D Quantum Gravity in terms of the matrix models, which seemed totally different from the conventional field theory but in the end was proven to be equivalent in the local limit.}, independent on peculiar mechanisms of energy pumping nor the boundary conditions as long as this energy pumping is provided.

In the WKB limit the tails of the PDF for velocity circulation $\Gamma$ over large fixed loops $C$ are controlled by a classical field $\phi_a^{cl}(r)$ (instanton) concentrated around the minimal surface bounded by $C$.

The  field is discontinuous across the minimal surface which leads to the delta function term for the tangent components of vorticity as a function of normal coordinate. 

The flux is still determined by the normal component of vorticity, which is smooth.

\section{Energy Flow From the Uniform Forces}

The popular belief in the turbulent community (which I share as well) is that the energy is pumped into the turbulent flow from the largest spatial scales, and dissipated at the smallest scales due to viscosity effects after propagating in the so called inertial range. 

Let us see how that happens in some detail. Using \NS{} equation \eqref{NSv} with constant uniform force $\vec f$ absorbed into the pressure as a boundary condition
\begin{eqnarray}
    && p(\vec r \ra \infty) \ra -f_\alpha \ral;
\end{eqnarray}
we have for the energy derivative

\begin{eqnarray}
   && \d_t \int d^3 r \oh \val^2 = \nonumber \\
   &&\int d^3 r \left( \nu \val \dbe^2 \val   - \val \left(\vbe \d_\beta \val + \d_\alpha p\right)\right)
\end{eqnarray}

Integrating by parts using the Stokes theorem we reduce this to
two expressions of the energy flow (dissipated equals incoming)
\begin{eqnarray}
    &&\mathcal{E} = \nu \int_V d^3 r \oal^2 =\nonumber\\
    &&\int_{\d V} d \sigma_\beta  \left(\vbe \left( p + \oh \val^2\right) +\nu \val e_{\alpha\beta\gamma} \oga\right) 
    \label{BoundaryFlow}
\end{eqnarray}

Velocity is related to vorticity by the Biot-Savart law \eqref{BS} with implied boundary condition of vanishing velocity at infinity.
In that case there is only one term contributing to the flow through the infinite sphere: the term $-f_\alpha \ral$ in the pressure.
This term can be reduced back to the usual volume integral of velocity times force
\begin{eqnarray}
    &&\mathcal{E} = f_\alpha Q_\alpha(\vec f);\\
    && Q_\alpha(\vec f) = \int_V d^3 r \val
    \label{QV}
\end{eqnarray}
Note that this asymptotic flow is laminar and purely potential, as vorticity is located far away from the boundary\footnote{
This geometry, with finite cell confining vorticity and energy flow being pumped from a distant boundary surface, was recently realized in beautiful experiments\cite{IM19}, where the vortex rings were initially shot from the eight corners of a glass cubic tank, and a stable vorticity cell (a confined vorticity blob in their terms) was created and observed and studied in the center of the tank.
The energy was pumped in pulses from eight corners and the vorticity distribution inside the cell was consistent with K41 scaling.
Reynolds numbers in that experiment were not large enough for our instanton, but at least the energy flow entering from the boundary and dissipating in a vortex cell inside was implemented and studied in real water.
}.

This is not a realistic boundary condition, but neither are conventional random forces with some arbitrary long-wavelength support in Fourier space. This is just the simplest way to provide steady energy flow in the Hopf equation. The resulting turbulent blob in the bulk is supposed to be universal in the limit of vanishing force.

This net velocity $Q_\alpha(\vec f)$ depends of the constant uniform random force which is hidden in the boundary condition for pressure. In general, to find net velocity, one has to solve the steady equation in the whole domain including the inner region where vorticity is present. This constant uniform force influences the equilibrium distribution velocity and vorticity in the steady state, thus affecting the net velocity.
Surely, mean value of net velocity is zero, due to the symmetry of the Gaussian distribution of random force.

Computing this vector $Q_\alpha(\vec f)$ for arbitrary force is a hard problem in general, but as we shall see, this force tends to zero in the turbulent limit, so that we can keep only linear term in net velocity, which leads to calculable distribution of velocity circulation.

If we assume that net vorticity is zero and that vorticity is distributed in the finite region inside the fluid, we can use the asymptotic form of the Biot-Savart integral at infinity in \eqref{BoundaryFlow} after which this net velocity can be expressed in terms of vorticity distribution
\begin{eqnarray}
   Q_\alpha(f) \propto e_{\alpha\beta\gamma} \int d^3 r \rbe\oga
\end{eqnarray}
One could recover original form \eqref{QV} by integration by parts using $\oga = e_{\gamma\lambda\rho} \dla v_\rho$. 

\section{Clebsch Parametrization of Vorticity}

Let us go deeper into the hydrodynamics.

We parameterize the vorticity by two-component \CL{} field $\phi = (\phi_1,\phi_2) $:
\begin{equation}
    \oal =\oh e_{\alpha\beta\gamma}e_{i j} \dbe \phi_i \dga \phi_j
\end{equation}
The metric and topology of the \CL{} target space remains unspecified at this point.

The Euler equations are then equivalent to passive convection of the \CL{} field by the velocity field:
\begin{eqnarray}\label{CLEq}
     &&\d_t \phi_a = -\val \dal \phi_a\\
     &&\val(r) =\oh e_{i j}\left(\phi_i\dal \phi_j\right)^\perp \label{Vperp}
\end{eqnarray}
Here $V^\perp$ denotes projection to the transverse direction in Fourier space, or:
\begin{equation}
    V^\perp_\alpha(r) = V_\alpha(r) + \dal \dbe \int d^3 r' \frac{V_\beta(r')}{4 \pi |r-r'|}
\end{equation}
One may check that projection (\ref{Vperp}) is equivalent to the Biot-Savart law (\ref{BS}).

The conventional Euler equations for vorticity:
\begin{equation}
    \d_t \oal = \obe \dbe \val - \vbe \dbe \oal\label{EulerO}
\end{equation}
follow from these equations\footnote{We are going to work with Euler equations in the next sections until we shall study the corrections (viscosity anomalies) coming from the dissipation term}.

In \NS{} equations the \CL{} variables can still be used to parametrize vorticity\cite{SCH16}, though the equation of motion is no longer a Hamiltonian type. In fact, this equation is nonlocal, so it is not very useful. 

The reader can find details in original paper, here we just present this equation in our notations
\begin{subequations}
\begin{eqnarray}\label{NSCl}
    &&\d_t \phi_a = \nu \d^2\phi_a  - V_\alpha \dal \phi_a;\\
    && V_\alpha = \val + \frac{e_{\alpha\beta\gamma}\obe (\dga B -A_\gamma)}{\vec \omega^2};\\
    && A_\alpha = \nu e_{i j} \dbe \phi_j \dbe \dal \phi_i;\\
    && \oal (\dal B - A_\alpha)= 0
\end{eqnarray}
\end{subequations}
There are no time derivatives of the auxilliary field $B$, so it is supposed to be expressed in terms of instant value of $\phi$ from the last equation, using line integrals along vorticity lines  $\d_t \vec r = \vec \omega(r)$.
In the Euler limit $\nu \ra 0$ this vector $A_\alpha$ goes to zero, and so does the auxiliary field $B$, after which we are left with just an advection term.

The \CL{} field maps $R_3$ to whatever space this field belongs and the velocity circulation around the loop $C \in R_3$:
\begin{equation}
    \Gamma(C) = \oint_{C}  d \ral \val = \oint_{\gamma_2} \phi_1 d \phi_2 = \mbox{Area}(\gamma_2)
\end{equation}
becomes the oriented area inside the planar loop $\gamma_2 = \phi(C)$. We discuss this relation later when we build the \CL{} instanton.

The most important property of the \CL{} fields is that they represent a $p,q$ pair in this generalized Hamiltonian dynamics. The phase-space volume element $D \phi = \prod_x d \phi_1(x) d \phi_2(x)$ is invariant with respect to time evolution, as required by the Liouville theorem. We will use it as a base of our distribution.

The generalized \BL{} flow (\GBF{}) corresponding to stationary vorticity is described by $G_\alpha(x)=0$.
These three conditions are in fact degenerate, as $\dal G_\alpha =0$. So, there are only two independent conditions, the same number as the number of local \CL{} degrees of freedom. However, as we see below, relation between vorticity and \CL{} field is not invertible. 

We are going to neglect the viscosity term when establishing the singular instanton solution, but later we take this term into account and we find the viscosity anomaly (finite limit at $\nu \ra 0$). This anomaly leads to smearing the singularities, however, as we shall see in extreme turbulent limit $\nu \ra 0$ at fixed energy flow the viscosity term disappears and Euler singularities reappear.

\section{Gauge invariance}

There is some gauge invariance (canonical transformation in terms of Hamiltonian system, or area preserving diffeomorphisms geometrically)\footnote{I am grateful to Pavel Wiegmann for drawing my attention to this invariance.}. 

\begin{eqnarray}\label{SDiff}
    &&\phi_a(r) \Ra M_a(\phi(r))\\
    &&\det \pbyp{M_a}{\phi_b} = \frac{\d(M_1,M_2)}{\d(\phi_1,\phi_2)} = 1.
\end{eqnarray}
These transformations manifestly preserve vorticity and therefore velocity. 

These variables and their ambiquity were known for centuries\cite{Lamb45} but they were not utilyzed within hydrodynamics until pineering work of Khalatnikov\cite{khalat52}.

Later, in the papers of Kuznetzov and Mikhailov\cite{KM80} and Levich\cite{L81} in early 80-ties, the topological meaning of the \CL{} variables was discovered and utilised. Modern mathematical formulation in terms of \SYM{} was initiated in\cite{M83}. 

Derivation of K41 spectrum in weak turbulence using  kinetic equations in \CL{} variables was done by Yakhot and Zakharov\cite{YZ93}, without referring to their topology nor the gauge invariance. 

In my old work\cite{TSVS} the \CL{} variables were identified as major degrees of freedom in statistics of vortex cells and their potential relations to string theory was suggested. 

Then, in recent work\cite{M19e} I suggested that the surface degrees of freedom of the vortex cells as $U(1)$ compactified critical $c=1$ string in two dimension, which was exactly solved  by means of matrix models. 

These were all the blind steps in the right direction, as I see it now.

In terms of field theory, this \SYM{} symmetry is an exact gauge invariance, rather than the symmetry of observables, much like color gauge symmetry in QCD. This is why back in the early 90-ties I referred to \CL{} fields as "quarks of turbulence". To be more precise, they are both quarks and gauge fields at the same time. 

It may be confusing that there is another gauge invariance in fluid dynamics, namely the $\bf{volume}$ preserving diffeomorphisms of Lagrange dynamics. Due to incompressibility, the volume element of the fluid, while moved by the velocity field, preserved its volume. 

However, these diffeomorphisms are not the symmetry of the Euler dynamics, unlike the $\bf{area}$ preserving diffeomorphisms of the Euler dynamics in \CL{} variables.

The space where the \CL{} fields belong to is not specified by their definition. For our theory it is important that this space is compact, which leads to discrete winding numbers. We accept the $S_2$ definition\cite{KM80,L81} 
\begin{equation}\label{S2param}
    \oal = \oh Z e_{i j k} e_{\alpha\beta\gamma} S_i \dbe S_j \dga S_k;\; S_i^2=1
\end{equation}
It can be rewritten in terms of our \CL{} fields using polar coordinates $\theta \in (0,\pi),\varphi\in (0,2\pi)$ for the unit vector $S=(\sin \theta \cos \phi, \sin \theta \sin \phi, \cos \theta)$:
\begin{eqnarray}\label{ClS2}
    &&\phi_1 = Z (1-\cos\theta); \\
    &&\phi_2 = \varphi \Mod{2 \pi}
\end{eqnarray}

The second variable $\phi_2$ is multi-valued, but vorticity is finite and continuous everywhere. The helicity $\int d^3 r \val \oal$ was ultimately related to winding number of that second \CL{} field \footnote{To be more precise, it was Hopf invariant on a sphere $S_3$ instead of real space $R_3$ (see\cite{KM80} for details).}.

The volume element on $S_2$
\begin{equation}
    d^2 \phi =  d \cos \theta d \varphi
\end{equation}
is equivalent to $d\phi_1 d\phi_2$ up to the scale factor $Z$.

From the point of view of the \SYM{} using the sphere as a target space for \CL{} field  amounts to \textbf{gauge fixing}, as we shall see in subsequent sections.

We are going to work in this gauge, where $\phi_2$ is an angular variable, as this will be the simplest one for topological properties.

One can introduce more general fluid dynamics with more than one \CL{} field and/or with higher genus of the \CL{} space, but we follow the Occam's razor here and stick to just one \CL{} field on a sphere without handles.

Note also that in the Euler dynamics our condition $G_\alpha=0$ comes from the Poisson bracket with Hamiltonian $H =\int d^3 r \oh \val^2$
\begin{eqnarray}\label{PBforGBF}
    && G_\alpha(r) = \PBR{\oal,H} =\int d^3 r' \fbyf{\oal(r)}{\phi_i(r')} e_{i j} \fbyf{H}{\phi_j(r')}=\nonumber\\
    &&-\int d^3 r' \fbyf{\oal(r)}{\phi_i(r')} v_\lambda(r') \dla{} \phi_i(r')
\end{eqnarray}
We only demand that this integral vanish. The stationary solution for \CL{} would mean that the integrand vanishes locally, which is too strong. We could not find any finite stationary solution for \CL{} field even in the limit of large circulation over large loop.  

The \GBF{} does not correspond to stationary \CL{} field: the more general equation
\begin{eqnarray}
    &&\d_t \oal = \int d^3 r' \fbyf{\oal(r)}{\phi_i(r')} \d_t \phi_i(r)\\
    &&\d_t \phi_i = -\val \dal \phi_i + e_{i j} \pbyp{h(\phi)}{\phi_j}  \label{GaugeTranClebsch}
\end{eqnarray}
with some unknown function $h(\phi)$ would still provide the \GBF{}.
The last term drops from here in virtue of infinitesimal gauge transformation $\delta \phi_a = \epsilon e_{a b} \pbyp{h(\phi)}{\phi_b}$ which leave vorticity invariant.

This means that \CL{} field is being gauge transformed while convected by the flow. For the vorticity this means the same \GBF{}.

\section{Invariant measure on GBF}

We now scale the factor $Z$ out of \CL{} field, the vorticity, velocity and net velocity
\begin{eqnarray}
    && \phi_1 \Ra Z \phi_1;\\
    &&\phi_2 \ra \phi_2;\\
   && \oal \Ra Z \oal;\\
   && \val \Ra Z \val;\\
   && Q_\alpha \Ra Z Q_\alpha\\
   &&G_\alpha \Ra Z^2 G_\alpha
\end{eqnarray}
after which $Z$ becomes a global variable, in addition to velocity and vorticity, which are determined by \CL field on a unit sphere $S_2$. It will be found later from the energy balance condition.

We propose the following invariant measure on \GBF{} manifold $\mathcal{G}$ parametrized by unit vector \CL field $\phi$, random force $\vec f$  and global parameter $Z$:
\begin{eqnarray}
    &&\int d \mu(\mathcal{G}) = \int d P(\vec f) d Z  D \phi D U D \Psi \nonumber \\
    &&\EXP{\i \sum_I U_I G^I + \oh \PBR{\Phi,\Phi}};\\
    && \Phi = \sum_I \Psi_I G^I
\end{eqnarray}
In addition to the original \CL{} field we have Lagrange multiplier field $U_\alpha(r)$ and the ghost Grassmann field $\Psi_\alpha(r)$, needed to compensate for the non-linearity of constraints.

We are using matrix notation where the spatial coordinate $r$ is treated as part of an index $I = (\vec r, \alpha), G^I = G_\alpha(\vec r) $ etc. The spatial integrals become sums and functional measure becomes product over space of local measures. 

The \PB of the $\Phi$ with itself does not vanish because this is Grassmann functional: integral of the Grassmann $\Psi_I$ field over space. The antisymmetric \PB of the Bosonic field $G^I$ matches the anti-commutation of $\Psi_I$ to produce non-vanishing \PB{}.
This measure is manifestly gauge invariant due to gauge invariance of the \PB{} as well as linear phase space $D\phi$.

There is a hidden supersymmetry in this measure which becomes manifest if we introduce a superfield
\begin{eqnarray}\label{SUSY}
    &&\mathcal{X}_\alpha(\vec r,\theta) = \Psi_\alpha(\vec r) + \theta U_\alpha(\vec r);\\
    &&D U D \Psi \EXP{\i \sum_I U_I G^I + \oh \PBR{\Phi,\Phi}} =\nonumber \\
    && D\mathcal{X} \EXP{ \int d \theta \left(\i \mathcal{P}(\theta) + \oh \theta \PBR{\mathcal{P}(\theta), \mathcal{P}(\theta)}\right)} ;\\
    && \mathcal{P}(\theta) =  \int_r  \mathcal{X}_\alpha(\vec r,\theta) G_\alpha(\vec r)
\end{eqnarray}
The Grassmann shift (or \mbox{BRST} transformation)
\begin{eqnarray}
    && \delta \theta = \epsilon;\\
    && \delta \Psi_\alpha = -\epsilon  U_\alpha;\\
    && \delta U_\alpha = 0;
\end{eqnarray}
leaves the superfield invariant.

Let us prove (to a physicist) that this phase space measure covers our manifold $\mathcal{G}$ uniformly.

The integral over the vector field $U_{r,\alpha}$ projects on  $\mathcal{G}$ , so that only linear vicinity of this hyper-surface in phase space contributes
\begin{equation}
    L: {\phi_{x,a} = \phi^\star_{x,a} + \Xi_{x,a}}
\end{equation}

In this linear vicinity we have Gaussian integral
\begin{eqnarray}
    &&\int_L D\mathcal{Z} = \int D\Xi D U D \Psi \nonumber\\
    &&\EXP{\i \sum_{I P}  U_I G^{I P} \Xi_P + \oh \sum_{P Q} Y^P E_{P Q} Y^Q};\\
    && Y^P = \sum_I \Psi_I G^{I P};\\
    && P = (\vec r, a), Q = (\vec r', b); a,b = 1,2;\\
    && E_{P Q} = \delta_{\vec r \vec r'} e_{a,b} ;\\
    && G^{I P}  = \pbyp{G^I[\phi^\star]}{\phi_P}
\end{eqnarray}

This integral involves the matrix $G^{I P}$ which so far depends upon the point $\phi^\star$ on a \GBF{} hyper-surface. Let us prove that this dependence cancels out.

We use so called SVD\cite{SVD}, well known in mathematics but rarely used in theoretical physics.
\begin{equation}
    G^{I P} = \sum_\lambda W_\lambda^I g_\lambda V_\lambda^P
\end{equation}
where $W, V$ are orthogonal matrices in their corresponding spaces\footnote{one of these matrices is not fully represented in this sum, as the number of singular values is bounded by the smallest of the ranks of $W,V$.}.

It is important however, that the dimensions of these two spaces are different : $ W \in O(3 N), V \in O(2 N)$, where $N$ is a number of points in space used to approximate the operator by a matrix. 

In this case there there are $2N$ or less positive eigenvalues $g_\lambda$ corresponding to square roots of eigenvalues of symmetric matrix $g^{P Q} = \sum_I G^{I P}G^{I Q}$ and the rest of eigenvalues are equal to zero.

This matrix is nothing but an induced metric on \GBF{} hyper-surface from embedding Hilbert space with Euclidean metric (see Appendix A. for a finite dimensional example).

In fact, there are some more zero modes with that metric, corresponding to the gauge invariance of the \CL{} representation:
\begin{equation}
    \sum_P G^{I P} \left(\delta_{gauge}\phi^{\star}\right)_P =0
\end{equation}

Obviously, only non-zero eigenvalues contribute to $G^{I P}$. We now perform orthogonal transformation in the variables $U,\Psi,\Xi$ absorbing corresponding matrices $W,V$. The linear measure $D\Xi D U D \Psi $ does not change, and we are left with sums over finite eigenvalues in exponential
\begin{subequations}
\begin{eqnarray}
    &&\i \sum_{\lambda}  U_\lambda g_\lambda \Xi_\lambda + \oh \sum_{\lambda \lambda'} Y_\lambda \hat{E}_{\lambda \lambda'}  Y_\lambda';\\
    &&Y_\lambda = \Psi_\lambda g_\lambda;\\
    &&\Psi_\lambda = \sum_I W_\lambda^I \Psi_I;\\
    &&U_\lambda = \sum_I W_\lambda^I U_I;\\
    &&\Xi_\lambda = \sum_P V_\lambda^P \Xi_P;\\
    &&\hat{E}_{\lambda \lambda'} = \sum_{P Q} V_\lambda^P V_{\lambda'}^Q E_{P Q}
\end{eqnarray}
\end{subequations}

Note that our matrix $V_\lambda^P$ is orthogonal but it it does not belong to symplectic group, so it does not leave invariant $E_{P Q}$. This symplectic symmetry of the Poisson brackets is related to Hamiltonian structure, which is not present in the \NS{} equation.

The linear measure 
\begin{equation}
    D\Xi D U D \Psi = \prod_{\lambda:g_\lambda \neq 0} d\Xi_\lambda d U_\lambda d \Psi_\lambda D\Omega 
\end{equation}
where $D\Omega $ is the volume element associated with zero modes (both for vector fields $U,\Psi$ and \CL fields $\Xi$).

Leaving the zero modes aside we can scale out the non-zero eigenvalues
\begin{subequations}
\begin{eqnarray}
    &&\Psi_\lambda \Ra \Psi_\lambda/g_\lambda;\\
    && U_\lambda \Ra U_\lambda/g_\lambda;\\
\end{eqnarray}
\end{subequations}
These eigenvalues then cancel in the measure by corresponding Jacobians
\begin{subequations}
\begin{eqnarray}
&&d \Psi_\lambda \Ra g_\lambda d \Psi_\lambda;\\
&&d U_\lambda \Ra \frac{1}{g_\lambda} d U_\lambda;
\end{eqnarray}
\end{subequations}

\section{Zero Modes and Gauge Fixing}

As we have mentioned already, this spherical parametrization is equivalent to gauge fixing. Let us discuss this in more detail.

Geometrically, the initial linear measure $d\phi_1 d\phi_2$ in phase space does not yet specify the metric of the space where $\phi$ belongs. The gauge transformations are the area preserving diffeomorphisms which change the metric tensor of the two dimensional space without changing its determinant.

Locally, two coordinates $\theta, \varphi$ correspond to the metric 
\begin{equation}\label{metric}
    d s^2 = d \theta^2 + \sin^2 \theta d\varphi^2
\end{equation} 
The measure $\sin\theta d \theta d \varphi = d (1-\cos \theta) d \varphi$ is linear in terms of \CL{} variables but the space is curved.

So, by specifying the $S_2$ metric in \CL{} space we fixed the gauge and substituted gauge symmetry with the $SO(3)$.

Now our field is the same as in well known sigma model, specifically it is $n=3$ n-field. The target space is now compact, with fixed area $4 \pi$. 

The Poisson brackets are replaced with
\begin{eqnarray}
    \PBR{F,G} \Ra  \int_r \fbyf{F}{S_j(\vec r)}e_{i j k} S_i(\vec r) \fbyf{G}{S_k(\vec r)}
\end{eqnarray}
The crucial difference between this theory and the sigma model is that the Lagrangean of the sigma model was only invariant with respect to $O(3)$ rotations of $\vec S$, but in our case there is a hidden gauge symmetry, changing the metric of the target space while preserving its topology and its area.

This hidden gauge symmetry comes about because effective Lagrangean only depends of the vorticity, which allows to change the metric in \CL{} space. So, there is a nontrivial mathematical problem\cite{Nikita} of description of the gauge orbit in functional space of all two-dimensional metrics. 

This problem, however, is \textbf{global} rather than local. We do not have independent \SYM{} in every point in space, we rather have one gauge orbit intersected by a single gauge condition (spherical metric). The gauge fixing takes place in target space rather than a physical times target space like in gauge theories.

The computation of determinants for non-zero modes in the previous section proceed in the same way, with an obvious modification. The two-dimensional field $\Xi(\vec r)$ now correspond to two coordinates in the tangent plane to the sphere $S_2$ at the particular \GBF{} 
\begin{eqnarray}
    &&\vec S = \vec S^\star + \vec e_1 \Xi_1 + \vec e_2 \Xi_2; \\
    && (\vec e_1\vec S^\star)  = (\vec e_2\vec S^\star)  = 0;\\
    && (\vec e_i \vec e_j) = \delta_{i j};\\
    && d^2 S = d \Xi_1 d \Xi_2
\end{eqnarray}

Repeating the steps of the integration over linear deviations from the steady state we find now after cancellation of nonzero modes
\begin{eqnarray}
&& \int_L D\mu \propto \int d Z d \Omega
\end{eqnarray}

Now we are prepared to fix the gauges. There are two gauge conditions here. The trivial one corresponds to a zero mode $U_\alpha^0 = \dal f $ with some scalar function $f(\vec r)$ vanishing at infinity. 

The standard linear gauge condition
\begin{equation}
    \dal U_\alpha =0
\end{equation}
leads to Jacobian of Laplace operator $\d^2 f$ and does not produce any dependence of remaining dynamic variables.

The nontrivial gauge fixing of \CL{} field is discussed in Appendix C.

We conclude that with the spherical representation as a \CL{} gauge condition, and $\dal U_\alpha =0$ as gauge condition for Lagrange multiplier field $U_\alpha$ the measure is uniform over the \GBF{} space.

We only considered obvious zero modes, related to \SYM{} and incompressibility. There are some other, hidden zero modes, which make our \GBF{} space much richer. 

These hidden zero modes are surfaces of vorticity singularities, which are responsible for multi-fractal scaling of Turbulence as well as the Area law in our theory.

\section{\CL{} instanton}

We found in\cite{M20a}  multi-valued fields with nontrivial topology which are relevant to large circulation asymptotic behavior.

In the following subsections we describe this instanton solution in some detail and discuss its topology and its physical properties.

We neglect viscosity which will be justified later when we find out that viscosity leads to smearing of singularities at some scale $h$ which tends to zero together with $\nu$ in  the turbulent limit. Until that we are going to work with Euler equations.

\subsection{Gauge Invariance and Clebsch Confinement}

The Turbulence phenomenon in fluid dynamics in \CL{} variables resembles the color confinement in QCD.

We have no Yang-Mills gauge field here, but instead we have nonlinear \CL{} field participating in gauge transformations. These transformations are global as opposed to local gauge transformations in QCD, but the common part is that this symmetry stays unbroken and leads to confinement of \CL{} field.

The description of \CL{} field as nonlinear waves\cite{YZ93} which was appropriate at large viscosity, or weak turbulence, quickly gets hopelessly complex when one tries to go beyond the K41 law into fully developed turbulence. The basic assumption\cite{YZ93} of the Gaussian distribution of \CL{} field breaks down at small viscosity.

The small viscosity in \NS{} equations is a nonperturbative limit, like the infra-red phenomena in QCD, when the waves combine into non-local and nonlinear structures best described as solitons or instantons.

Nobody managed to explain color confinement in gauge theories as a result of strong interaction of gluon waves. On the contrary, the topologically nontrivial field configurations such as monopoles in 3D gauge theory and instantons in 4D led to the understanding of the color confinement.

This is what we are doing here as well, except our singular solutions are not point like singularities but rather singular vorticity sheets.

Vorticity sheets (so called Zeldovich pancakes\cite{SZ89}), were extensively discussed in the literature in the context of the cosmic turbulence. Superficially they look similar to my instantons but at closer look there are some important distinctions. For one thing they are unrelated to the random surfaces, and for another one, they seem to have no topological numbers.

The general physics of the "frozen" vorticity in incompressible flow, collapsing in the normal direction and expanding along the surface, is essentially the same. What is different here is an explicit singular solution with its tangent and normal components at the surface, the \CL{} field topology and its consequences for the circulation PDF.

The relevance of classical solutions in nonlinear stochastic equations to the intermittency phenomena (tails of the PDF for observables) was noticed back in the 90-ties\cite{FKLM} when it was used\cite{GM96} to explain intermittency in Burgers equation.

However, nobody succeeded in finding the instanton solution in 3D fluid dynamics until now.
Remarkably, though, our instantons are three-dimensional analogs of the shocks in one-dimensional Burgers turbulence.

\subsection{Clebsch \DS{}}

Let us now describe the proposed stationary solutions of Euler equations in \CL{} variables.

Our \CL{} field $\phi_2$ has $2 \pi n$ discontinuity across some surface $S_C$ bounded by $C$. As it is argued in  previous papers\cite{M20a,M20b,M20c} the minimal surface is compatible with \CL{} parametrization of conserved vorticity directed at its normal in linear vicinity of the surface.

In general case a minimal surface can be described by Enneper-Weierstrass parametrization\cite{EW}:
 \begin{eqnarray}\label{Enneper}
    &&\vec X(\rho,\theta) = \Re \vec F\left(\rho e^{\i \theta}\right);\\
    &&\vec F'(z) = \left\{\oh(1-g^2)f, \frac{\i}{2} (1+g^2)f, g f\right\} 
 \end{eqnarray}
 with $g(z), f(z)$ being analytic functions inside the unit circle $|z| <1$.
 Such surface is shown at Fig.\ref{fig::Enneper} for $f = 1, g = z$:
 \begin{figure}
    \centering
    \includegraphics[width=0.9\textwidth]{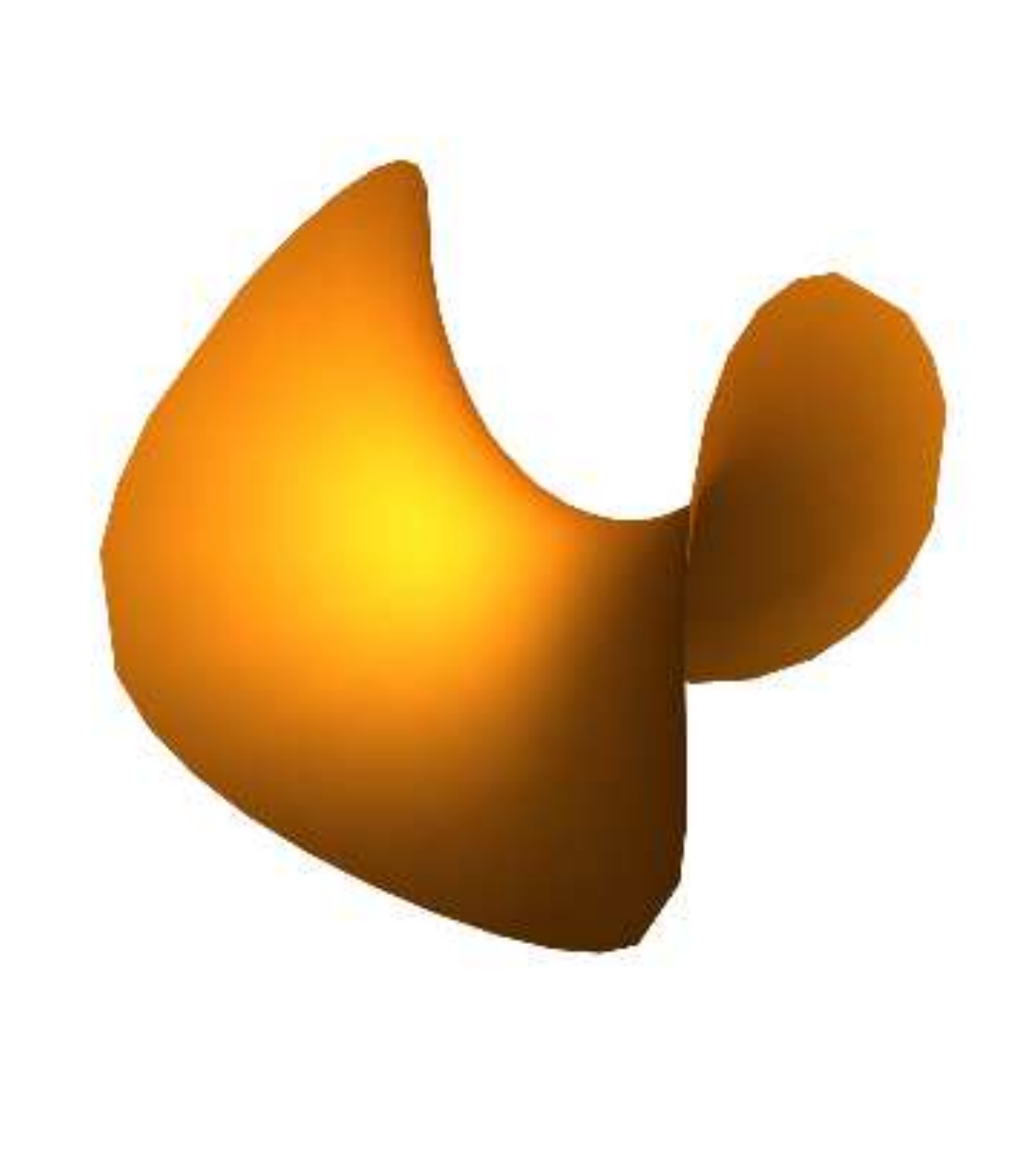}
    \caption{The Enneper's Minimal surface with $f = 1, g = z$.}
    \label{fig::Enneper} 
\end{figure}

However, there exist stationary solutions of the Euler equation with \CL{} discontinuity at arbitrary, non-minimal surface. \footnote{The minimal surface will presumably dominate at large loop, because the effective string tension for the random surface is large here unlike QCD -- of the order of $1/h^2$ where $h \ra \nu^{\frac{3}{5}}$ is an effective thickness of the surface, created by the viscosity in \NS{} equation, as we shall see below. }

Let us discuss this important point in some detail.
Let us assume that the \CL{} field is discontinuous across some generic smooth surface, bounded by the loop $C$ where we specify the circulation.  In that case the non-singular part of vorticity $\oal^{ns}$ will involve tangent derivatives of the \CL{} field, and therefore this $\oal^{ns}$  will be directed at the local normal to the surface.

In quadratic vicinity of local tangent plane to the surface its equation reads ( with $K_1, K_2$ being principal curvatures at this point)
\begin{subequations}
\label{ClebschAtS}
\begin{eqnarray}
  &&S(x,y,z) = z - \frac{K_1}{2} x^2 - \frac{K_2}{2} y^2 =0\\
  &&n_\alpha = \frac{(-K_1 x , -K_2 y, 1)}{\sqrt{1 + K_1^2 x^2 + K_2^2 y^2}} \ra (-K_1 x , -K_2 y, 1);\\
  &&\dal n_\alpha = -(K_1 + K_2)
\end{eqnarray}
\end{subequations}
The minimal surface would correspond to $K_1 + K_2 =0$, which means a conserved normal vector along the surface.
In general case this conservation is not required by the Euler equations.

As we shall see now, the Euler equations lead to restrictions of the normal derivatives $ n_\alpha\dal \phi_a$ of the \CL{} field.

It is easy to see that the Euler equations \eqref{GaugeTranClebsch} demand that the normal velocity $n_\alpha \val=0$ at the surface, to cancel the $\delta\left(S(x,y,z)\right)$ terms in convection term $\val \dal \phi_a$. The tangent derivatives $ (\dal- n_\alpha n_\beta \dbe) \phi_a$ are in general finite and can balance with the gauge transformation term.

Vanishing normal velocity means that the normal derivatives of \CL{} field at the surface, taken as a limit from each side, drop from the convection term $\val \dal \phi_a$ and therefore from the dynamics. 

These normal derivatives can be specified as initial conditions. The \CL{} field is frozen in the flow, sliding along the singular surface modulo gauge transformations. There is no flow in the normal direction at the surface, so the \CL{} field with $n_\alpha\dal \phi_a =0$ will be stationary. 

This boundary condition is manifestly gauge invariant, so it is preserved not only by convection term $\val \dal \phi_a$ but also by the second term, the gauge transformation. No other Neumann condition $n_\alpha\dal \phi_a = F_a(x,y) $ would be gauge invariant. 

Contrary to some of my early conjectures \cite{M20b, M20c}, there are no apparent restrictions from Euler dynamics on the \DS{}.
Solving the equation $n_\alpha\dal \phi_a =0$ in quadratic vicinity we find the linear term of Taylor expansion in $z$ for the non-singular part of \CL{} field $\phi^{ns}_a$
\begin{eqnarray}
   \phi_a^{ns}(x,y,z) = \phi_a^{ns}(x,y,0) + z \left(K_1 x\d_x + K_2 y \d_y\right) \phi_a^{ns}(x,y,0) + \dots
\end{eqnarray}
The non-singular part of vorticity $\omega^{ns}_\alpha = e_{\alpha\beta\gamma} \dbe \phi^{ns}_1\dga \phi^{ns}_2 $ is identically conserved in this linear vicinity, for arbitrary $K_1, K_2$. 

Note that this \DS{} is related to two surfaces of constant $\phi_1, \phi_2$ which are usually considered in geometric interpretations of the \CL{} field. 

These two surfaces are in fact both normal to our \DS{} at every point, as both \CL{} fields are constant in the normal direction. The normal vector of \DS{} is proportional to the cross product of these two normal vectors of constant \CL{} surfaces. 

One of these surfaces,  corresponding to $\phi_2 = \mbox{const}$, ends at our \DS{}, as the other side of the \DS{} corresponds to $2\pi n$- shifted constant value of $\phi_2$ in normal direction.

We parametrize the \DS{} as a mapping to $R_3$ from the unit disk in polar coordinates $\rho,\alpha$
\begin{equation}
    S_C: \vec r = \vec X(\xi),\; \xi = (\rho,\alpha)
\end{equation}

In the linear vicinity of the surface
\begin{equation}
    \delta S_C:\vec r = \vec X(\xi)+ \eta \vec n(\xi)
\end{equation}
the \CL{} field $\phi_2$ is discontinuous
\begin{equation}
   \phi_2\left(\vec r \in \delta S_C\right) = m \alpha + 2 \pi n \theta(\eta) + O(\eta^2);\; m,n \in \Z
\end{equation}
while the other component is continuous
\begin{equation}
    \phi_1\left(\vec r \in \delta S_C\right)   = \Phi(\xi) + O(\eta^2)
\end{equation}

The vorticity has the delta-function singularity at the surface:

\begin{subequations}
\begin{eqnarray}
    && g_{i j} = \d_i X_\mu(\xi) \d_j X_\mu(\xi);\\
    &&\vec \omega\left(r \in \delta S_C\right) \ra 
    \delta(\eta) 2\pi n \vec \nabla \Phi(\xi)\times \vec n(\xi) + \vec n(\xi) \Omega(\xi) ;\\
    &&\Omega(\xi) = \frac{m\pbyp{\Phi(\xi)}{\rho}}{\sqrt{\det g}};\\
    && \vec n = \frac{\d_\rho \vec X \times \d_\alpha \vec X}{\sqrt{\det{ g}}};
\end{eqnarray}
\end{subequations}

This delta term in vorticity is orthogonal to the normal vector to the surface and thus does not contribute to the flux through the minimal surface, so this flux is still determined by the second (regular) term and circulation is related to this $\Phi(\xi)$

\begin{eqnarray}\label{GammaPhi}
    &&\Gamma_C = \oint_C \val d \ral = \int_S d\phi_1 \wedge d \phi_2 =\nonumber\\
    &&m \int_{0}^{2\pi} \left(\Phi(1,\alpha) - \Phi(0,\alpha) \right)d\alpha
\end{eqnarray}

The Stokes theorem ensures that the flux through any other surface bounded by the loop $C$ would be the same, but in that case the singular tangent component of vorticity would also contribute. The simplest computation corresponds to choosing the flux through the \DS{}.

The instanton velocity field reduces to the surface integral
\begin{eqnarray}\label{BSvelocity}
    &&v^{inst}_\beta(r) =  2\pi n  \left(\delta_{\beta\gamma} \dal - \delta_{\alpha\beta} \dga\right) \nonumber\\
    &&\int_{S_C} d \sigma_\gamma(\xi)\dal \Phi(\xi) \frac{1}{4 \pi |\vec X(\xi)- \vec r|}
\end{eqnarray}

We are assuming that the \CL{} field falls off outside the surface so that vorticity is present only in an infinitesimal layer surrounding this surface. In this case only the delta function term contributes to the Biot-Savart integral though only a regular term contributes to the circulation.

Let us now consider the steady flow \CL{} equations derived in\cite{M20a} , which we call the master equation:
\begin{equation}\label{masterCL}
    \val \dal \phi_a = e_{a b} \pbyp{h(\phi)}{\phi_b}
\end{equation}
Here the gauge function $h(\phi)$ is arbitrary, and must be determined from consistency of the equation.

The master equation is much simpler than the vorticity equations for \GBF{}. 

The leading term in these equations near the \DS{} is the normal flow restriction 
\begin{equation}
    \val(r) n_\alpha(r) =0; r \in S_C
\end{equation}
which annihilates the $\delta(\eta)$ term on the left side of \eqref{masterCL}. 

The next order terms will already involve the gauge function $h(\phi)$. We found it the simplest to analyze the balance of singular terms directly in the \NS{} equation \eqref{GBFV} for velocity  (see section "Viscosity anomaly and Scaling Laws").

\subsection{\DS{} as zero mode and multi-fractals}

As we have seen in the previous section, there could be generic solutions of the Euler equations with discontinuity of the \CL{} field across arbitrary smooth surface. 

This makes the shape of this surface the zero mode of the \GBF{} measure in the Euler limit. In other words, we have to integrate over all such surfaces with some local measure.

Let us stress again, that the shape of the \DS{} is not fixed by the Euler equation, so it is conserved and is determined by initial conditions. The \CL{} field will flow with the fluid around these fixed surfaces, which remain steady, with the only condition that normal velocity as well as normal derivatives of the \CL{} field vanish at the \DS{}.

Our averaging of Hopf functional over initial (and boundary) conditions in the \GBF{} includes therefore averaging over \DS{} with arbitrary local invariant measure. It would remain the fixed point of the Hopf equation after averaging over these random surfaces, regardless of the measure.

The relation of turbulence statistics to random surfaces was conjectured in my old work \cite{TSVS}.

Let us reproduce these arguments here, with some new understanding we gained in the last 25 years. 

The simplest measure is well known Polyakov measure for random surfaces used in the noncritical string theory\cite{Pol81, KPZ}
\begin{eqnarray}\label{Pol}
&& d\mu(g,X) = D[g] D X \EXP{-\int_D d^2 \xi \sqrt{g} \left(\oh g^{i j}\d_i X_\alpha \d_j X_\alpha + \mu\right)};
\end{eqnarray}
with $g_{i j}$ being an internal metric on the surface.

The diffeomorphism invariance (reparametrization of coordinates $\xi = {\xi_1, \xi_2}$ on the surface) allows us to choose conformal metric $g_{i j}(\xi) = e^{\alpha \varphi(\xi)} \hat{g}_{i j}$, where $\hat{g}_{i j}$ is the base metric corresponding to the surface of fixed genus and boundary.

In conformal metric the Gaussian integral over $X_\alpha$ reduces to the exponential of the classical action (minimal surface area bounded by the loop $C$) divided by the square root of the determinant coming from fluctuations of $X$ field  around that minimal surface.

In case of free closed surface the minimal surface shrinks to a point. In case of fixed boundary (or some number of points pinned in space) there is a nontrivial minimal surface bounded by these points and loops.

The resulting effective action for the field $\varphi(\xi)$ is the Liouville theory \cite{Pol81,DK81}
\begin{eqnarray}
\label{Liouville}
&& d\mu(g) = D\varphi \EXP{-\frac{1}{4\pi} \int_D d^2 \xi \oh \hat{g}^{i j}\d_i \varphi \d_j \varphi - Q
\hat{R} \varphi + \mu e^{\alpha\varphi}};
\end{eqnarray}
where $ \hat{R} $ is the scalar curvature in the background metric $\hat{g}_{i j}$.

The parameters $ Q, \alpha $ should be found from the
self-consistency
requirements. In case of the ordinary string theory in $ d $ dimensional space the requirement of cancellation of conformal anomalies yields
\begin{eqnarray}
   &&\alpha = \frac{\sqrt{1-d} - \sqrt{25-d}}{2 \sqrt{3}};\\
    &&Q = \sqrt{\frac{25-d}{3}}
\end{eqnarray}

In three dimensions $ \alpha $ is a complex number, which is fatal for the string theory. Numerical simulations have shown that in fact the free random surfaces in three dimensions are not smooth -- they degenerate into branched polymers.

Fortunately this formula does not apply to
turbulence, because the dynamics of the $ X $ field is
different here.

This particular random surface is not completely free. Being the \DS{} it cannot intersect itself.  

In that sense it is similar\cite{M19e} to the phase boundaries in the 3D Ising model, which are known to be stable. Analogy is incomplete, because the discontinuities here are described by an arbitrary integer winding number $n$, unlike just one type of phase boundary in the Ising model.

Perhaps, the higher derivative terms, corresponding to "rigid random surface"\footnote{I am grateful to Nikita Nekrasov for this comment.} would prevent these self-intersections. The simplest such term in the exponential is the square of gradient of the normal vector
\begin{eqnarray}
   && \EXP{-\oh\int_D d^2 \xi \sqrt{\hat g} g^{i j}\d_i N_\alpha \d_j N_\alpha};\\
   &&N_\alpha = g^{i j}e_{\alpha\beta\gamma}\d_i X_\beta \d_j X_\gamma
\end{eqnarray}

This term introduces quartic interaction in the dynamics of the $\vec X$ field. 

In conformal metric this term will now explicitly couple the metric field $\varphi$ to the surface coordinate field $X$:
\begin{eqnarray}
    && g_{ i j} =e^{\alpha\varphi} \hat{g}_{i j};\\
   && \oh\int_D d^2 \xi \sqrt{\hat g} \hat{g}^{i j}\d_i N_\alpha \d_j N_\alpha;\\
   &&N_\alpha = e^{-\alpha\varphi} \hat{g }^{i j}e_{\alpha\beta\gamma}\d_i X_\beta \d_j X_\gamma
\end{eqnarray}
The square of gradient of the normal vector effectively prevents the surface from self-intersection as the normal vector jumps at this self-intersection.

Another difference is that for a closed surface the volume inside is a motion invariant in the Euler equation, as the normal velocity vanishes everywhere on our \DS{}. Therefore there is an extra restriction on fluctuating shape $X_\alpha$ in Euler dynamics
\begin{eqnarray}\label{Vol}
    \mbox{Vol}[X] = \frac{1}{6}e_{\alpha\beta\gamma} \oh e_{i j}\int_D d^2 \xi X_\alpha \d_i X_\beta  \d_j X_\gamma = \mbox{const};
\end{eqnarray}

In our statistics, the Euler-conserved quantities become exponentially distributed with some Lagrange multiplier, due to re-connection of these surfaces in full \NS{} dynamics. 

These conserved quantities for a subsystem under consideration are exchanged by viscous effects with the thermostat (remaining system), which leads to exponentiation of constraints due to the imaginary saddle point in the Fourier integral for the Lagrange multiplier (see Appendix B).

\begin{equation}
    \EXP{- \lambda \mbox{Vol}[X] }
\end{equation}
This term in effective action was suggested in my old work \cite{TSVS}.

In the lowest (second) order in derivatives we just have the quadratic Polyakov Action \eqref{Pol} plus this volume term.

This makes this closed random surface equivalent to fluctuating soap bubble (as opposed to soap film bounded by the "wire" $C$). Everybody with kids knows that soap bubbles as well as soap films exist in three dimensions :).

For the closed surface the second \CL{} field can be just constant on both sides of the bubble. We can choose this constant to be zero outside, then inside the bubble $\phi_2 = -2\pi n$. 

In this case there is no normal component of vorticity at the surface of the bubble, just the delta function $2 \pi n \delta(\eta) \nabla \phi_2(\xi_1, \xi_2) \times \vec n$ for the tangent components. The circulation around any loop at the surface of the bubble equals to zero. 

This vanishing circulation was conjectured in \cite{TSVS} based on different arguments: it was assumed that there was some vorticity inside the bubble. Our new theory is based on singular surfaces rather than three dimensional vorticity cells.

Anyway, the effective action for the remaining conformal metric field $\varphi$ will be now different from the above Liouville action.

The microscopic derivation of the parameters in the measure  $\mu, \alpha, Q$ would require full solution of the \NS{} equation in the viscous layer around the \DS{}. This is needed because the singularity is an idealization of the vorticity peak in Zeldovich pancake. 

In presence of viscosity these zero modes are replaced by an actual dynamics of the velocity when the smeared singularity surface bends and approaches self-intersection. The problem is no longer a dynamics of the two-dimensional surface, it is a dynamics of three dimensional velocity/vorticity field.

So far we only know that the thickness $h$ of this layer goes to zero  as some power of viscosity (see below). The values of $\mu,\alpha,Q$ remain unknown. 

At small loops $C$ in the units of string tension $\mu$, the surfaces will fluctuate strongly, and that could be the  source of multi-fractal scaling in turbulence.

We consider the surface pinned at several points, separated by a distance small compared to the mean size $\frac{1}{\sqrt{\mu_{eff}}}$ of the random surface area. 

One could obtain this surface in our loop Hopf functional by selecting the loop consisting of multiple small loops. 
This would lead to the the surface bounded by all these loops, with topology of a sphere with some number of holes.

\begin{figure}
    \centering
    \includegraphics[width=0.9\textwidth]{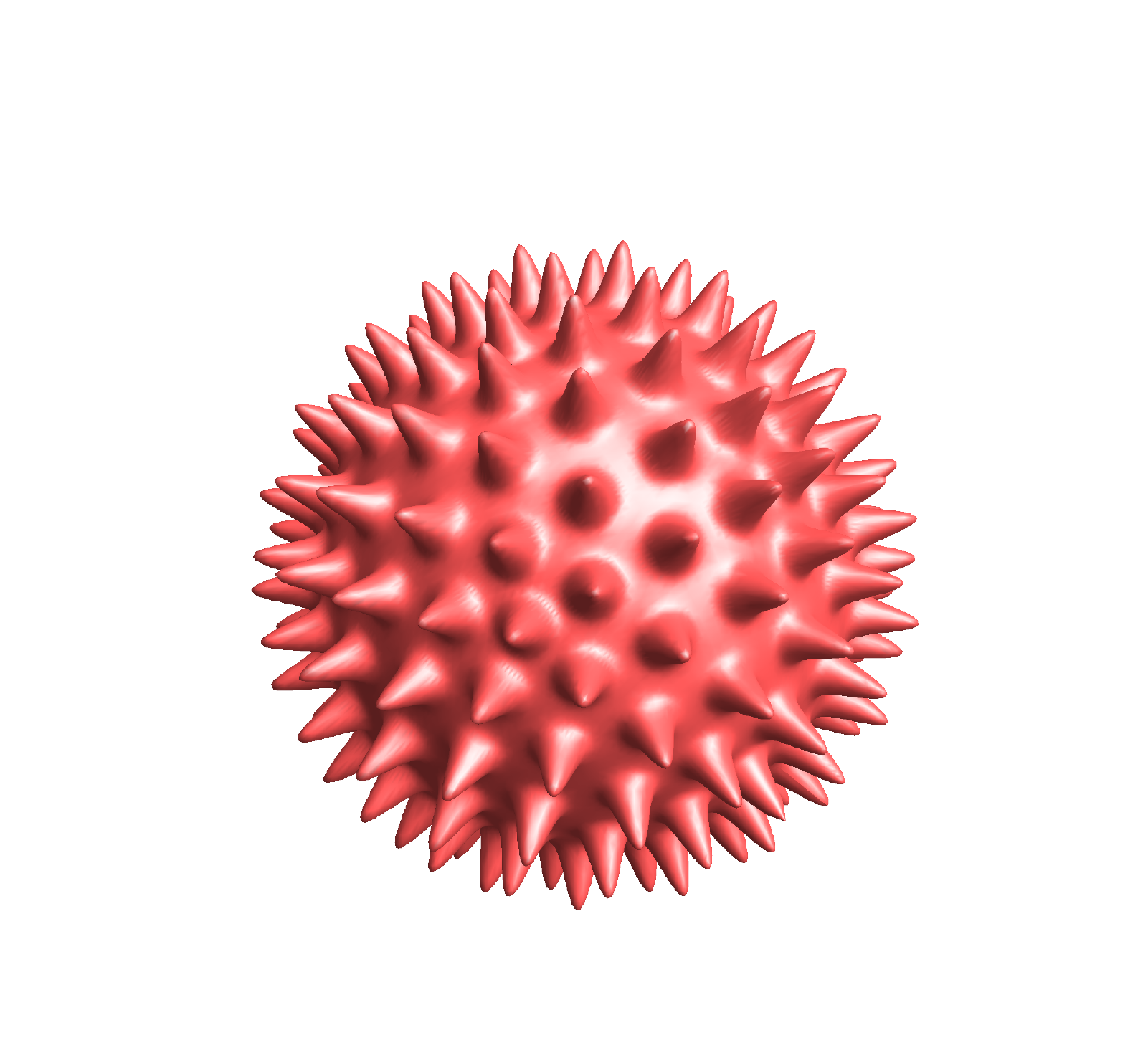}
    \caption{The sphere pinned to multiple points}
    \label{fig::SpikyBall} 
\end{figure}

In the limit of each of these loops shrinking to a point, it is equivalent to pinning the surface to multiple points in space, resembling COVID-19 virus.
Fig.\ref{fig::SpikyBall}.

What we get in the limit can be expressed in terms of  the
{\em vertex operator} of the string theory
\begin{equation}
 V(r) \propto\int_{S} d^2 \xi 
\sqrt{g}\delta^3(X(\xi) -r) ;\
\end{equation}
The important detail here is the factor $ \sqrt{g} \propto e^{\alpha \varphi} $ corresponding to the metric tensor at the surface.

The properties of such vertex operators were studied in the string theory\cite{DK81}.
The short distance correlation functions we need here, would all reduce to a Gaussian integral with logarithmic  correlation function. The surface tension $\mu$ can be neglected in this UV limit.

The moments of $ V $ would behave as
\begin{equation}
\VEV{V(\vec r_1)\dots V(\vec r_n)} \propto r_0^{-\Delta(n)}; \Delta(n) = \oh n\alpha(n\alpha+Q)
\end{equation}
where $|\vec r_i - \vec r_j| \sim r_0$ is an ultraviolet cutoff (viscous scale in our case). 

In our theory, the effective action is not the Liouville theory \eqref{Liouville}. Still, in the low gradient limit it has the same general form but with different parameters $Q, \alpha$. These parameters are to be found from the self-consistency conditions, just like they were found from requirement of cancellation of conformal anomalies in the string theory.

In case of the turbulence theory we have multi-fractal scaling laws \eqref{Zeta}, which in this case will involve vorticity $\vec \omega = \nabla \times \vec v$, as the limit of the circulation around infinitesimal loop. Assuming multi-scaling index $\zeta(n)$ just shifted by $1$ for each gradient we have
\begin{eqnarray}
   \VEV{V(\vec r_1)\dots V(\vec r_n)} \sim \VEV{\vec \omega(\vec r_1)\dots \vec\omega(\vec r_n)} \sim r_0^{\zeta(n) -n};
\end{eqnarray}
which implies that 
\begin{eqnarray}
   \zeta(n) = n -\oh n\alpha(n\alpha+Q)
\end{eqnarray}
We know exact value $\zeta(3) =1$, from which we may express $Q$
\begin{eqnarray}
&& Q = \frac{4-9 \alpha^2}{3 \alpha};\\
   &&\zeta(n) = \frac{n}{3} + \frac{ \alpha^2 n(3-n)}{2}
\end{eqnarray}
It is interesting that ideal K41 scaling would correspond to $\alpha \ra 0, Q \ra \frac{4}{3\alpha}\ra \infty$.

We fitted the data \cite{SK20} and the best fit
\begin{eqnarray}
   &&\alpha =  0.185293;\\
   && Q = 6.63995;
\end{eqnarray}

\begin{figure}
    \centering
    \includegraphics[width=0.9\textwidth]{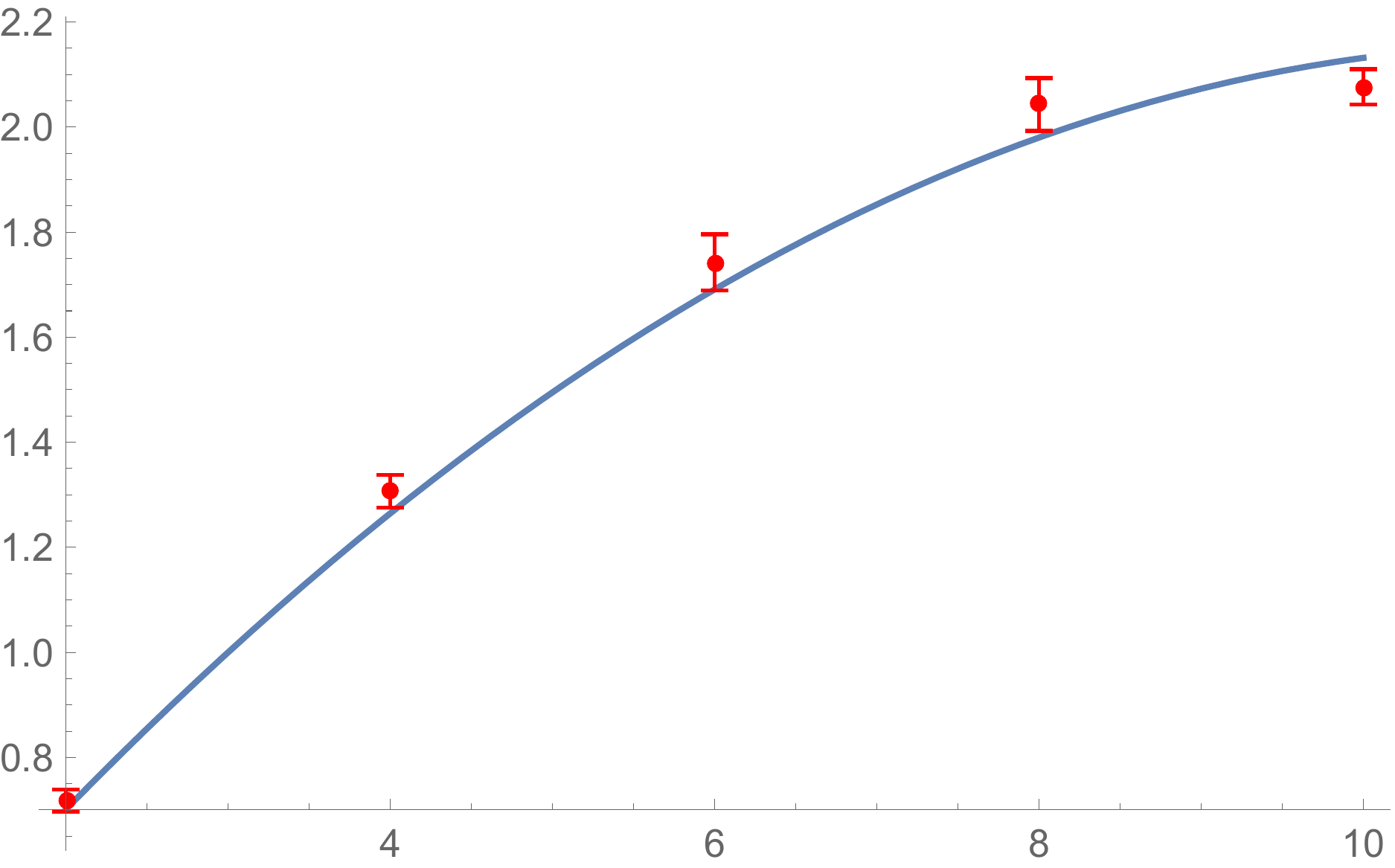}
    \caption{The model function $\zeta(n)$ with $\alpha = 0.185293$.}
    \label{fig::Zeta} 
\end{figure}

The plot of $\zeta(n)$ for this value of $\alpha$ is shown at Fig.\ref{fig::Zeta} together with error bars.
We used the data for transverse components of velocity, which is related to vorticity ($ \omega_z = \d_x v_y - \d_y v_x$ etc) we are in fact predicting.

We cannot claim this is an exact result for the multi-fractal dimensions, as we did not compute  $\alpha$ from the Navier-Stokes equation. All we can say is that Liouville Action with certain parameters can fit the existing data to some degree.

Our curve at large $n$ turns down and becomes negative, which is not what the DNS is telling us. So this curve may describe only small moments. 

At large $n$ the approximation neglecting the string tension $\mu$ in the Liouville Action no longer is valid, so the full Liouville theory must be used, as well as rigidity and volume conservation.

This missing microscopic computation, taking into consideration the volume conservation and rigidity of the \DS{} remains as an outstanding challenge. 

However, the higher moments of velocity circulation over the large loop, related to the same vorticity correlation functions integrated over the minimal surface inside the large loop, are calculable in our theory. 
\begin{eqnarray}
   \VEV{\Gamma_C^n} = &&\int_{S_C} d \vec \sigma(\vec r_1) \dots \int_{S_C} d \vec \sigma(\vec r_1) \VEV{\vec \omega(\vec r_1)\dots \vec\omega(\vec r_n)} \sim  R^{ n +\zeta(n)}
\end{eqnarray}

As we show below, these higher moments scale as $R^{n + \mbox{const}}$  where $R $ is a size of the loop. K41 law would correspond to $R^{\ft n}$, and multi-fractal scaling would correspond to $R^{ n +\zeta(n)}$. We fit the ratio of high moments as linear function of $R$. 

The constant term in $\zeta(n)$ cancels in the ratios of moments, so we see a perfect linear fit starting from $R_0 \sim 100\eta$ where $\eta$ is a Kolmogorov viscous scale, related to the energy flow density. This value of $R_0$ serves as an estimate of effective string tension
$\mu_{eff} \sim \frac{1}{R_0^2}$.

However, if you do log-log fit of the moments vs $R$ (rather than moments ratios) this constant addition could imitate shifted slope, and this can explain the power $R^{1.1 n}$ instead of $R^{n + \zeta(\infty)}$ which was found in \cite{S19} by log-log fit. This $\zeta(\infty) \sim 2$ is  positive, so it would imitate positive small shift $0.1 n$ of the fitted exponent, found in \cite{S19} for large $n \sim 10$.

So, if we interpret our asymptotic law $R^{n + \mbox{const}}$ as a large $n$ limit of multi-fractal law, this would mean that $\zeta(n) \ra \mbox{const}$ at large $n$, in agreement with direct measurements of $\zeta(n)$ in DNS.\footnote{As I learned from Kartik Iyer, he have also made this observation.}

At large loops the dominant surface would be the one with minimal area, in the same way as it happens with QCD string. There, too, the gluon field strength (analog of our vorticity) collapses to the surface of thickness small related to the size of the loop.

Note an important difference with the string theory. There, we were interested in the limit where the effective string tension $\mu_{eff}$ is much less than the UV cutoff in momentum space, because it was determining the physical mass spectrum.

Here, on the other hand, the limit of large loop will correspond simply to the loops larger than $R_0$. The \DS{} will reduce to a minimal surface in the whole  range of scales $ R > R_0$, usually associated with the strong turbulence. No need to assume large velocity circulation for that.

In the following we are going to treat the surface classically, assuming it coincides with the minimal surface.

\subsection{Instanton On Flat Surface}

Here we re-derive and correct the preliminary results described in the preprint\cite{M20b}. Some of the assumptions made in that paper turned out to be incorrect. The general predictions for PDF stay the same but formulas describing the dependence of the shape of the loop change significantly. 

The simplest case of our instanton is that of a flat loop in 3D space, which we assume to be in $x , y$ plane.
The minimal surface is a part $D_C$ of $x,y$ plane bounded by this flat loop.

The cylindrical coordinate system $\rho, \theta, z$ we are using has a fictitious singularity at the origin, where $\sqrt{ g} = \rho =0$. To keep the normal vorticity $ \omega_n \propto \frac{1}{\rho} \pbyp{\Phi}{\rho}$ finite at the origin the \CL{} field have to obey extra condition
\begin{equation}
    \d_i \Phi(\vec r = 0) = 0
\end{equation}
In other terms, the linear term of Taylor expansion of $\Phi$ at the origin must vanish otherwise the normal vorticity will have $1/|\vec r|$ pole.

The generic formula \eqref{BSvelocity} simplifies here (here $i,j = 1,2$):
\begin{subequations}
\begin{eqnarray}
    &&v^{inst}_i(r_0) \ra \pm \pi n \d_i \Phi ,\\
    &&v^{inst}_3(r_0) = \nonumber\\
    &&\frac{n}{2} \int_{D(C)} d^2 r \sqrt{g} g^{ i j} \d_i \Phi(r) \d_j \frac{1}{|r-r_0|}
\end{eqnarray}
\end{subequations}
The vanishing regular part of tangent velocity means that the regular part of equation \eqref{masterCL} is satisfied identically with $h=0$. 

As for the singular part, proportional to $\delta(z)$ it requires $v_z(r) =0$. 

In fact, there is always extra smooth contribution $\vec v^s(r_0)$ to the normal velocity from the 3D Biot-Savart integral of over vorticity in the remaining space (see\cite{M20a}). So, correct equation reads
\begin{eqnarray}
&&v_z(r_0) = v^s_z(r_0) + \nonumber\\
&&\frac{n}{2} \int_{D(C)} d^2 r \sqrt{g} g^{ i j} \d_i \Phi(r) \d_j \frac{1}{|r-r_0|} =0
\end{eqnarray}

\subsection{Minimization Problem}

There is a way to reduce our master equation to a minimization of a quadratic functional.

Let us  make the integral transformation 
\begin{equation}\label{PhiH}
    \Phi(\vec r) = \frac{\int_{D_C} d^2 r v^s_z(\vec r, z)}{n} \int_{D_C} d^2 r' \frac{H(\vec r')}{ 2 \pi|r - r'|}
\end{equation}
and we are arrive at universal equation 
\begin{eqnarray}
    &&\frac{1}{4 \pi^2}\int_{D_C} d^2 r'\dal \frac{1}{|\vec r' - \vec r|}  \nonumber\\
    &&\int_{D_C} d^2 r'' H(\vec r'')\dal'\frac{1}{ |r'' - r'|} = R(\vec r)
\end{eqnarray}

Here  
\begin{equation}
    R(\vec r) = \frac{v^s_z(\vec r, z)}{\int_{D_C} d^2 r v^s_z(\vec r, z)}
\end{equation}
is normalized to unit integral over the domain.

As we are interested in large size of domain $D_C$ compared to the size of vorticity support in the thermostat, this $R(\vec r)$ is concentrated inside a finite region near the center of $D_C$. Later we study this equation approximating $R(\vec r)$ by a delta function. Now we proceed for a general $R(\vec r)$.

We observe that this problem is equivalent to minimization of positive quadratic form
\begin{equation}\label{Target}
    Q[H] = -\int_{D_C} d^2 r H(r) R(\vec r) +  \oh\int_{D_C} d^2 r  F_\alpha^2[H,\vec r]
\end{equation}
where $\vec r_0$ is the center of the disk $D$
\begin{equation}\label{Falpha}
    F_\alpha[H,\vec r] = \frac{1}{2 \pi}\int_{D_C} d^2 r' H(\vec r') \dal' \frac{1}{|\vec r - \vec r'|} 
\end{equation}
As we shall see later, the position of the origin drops from asymptotic formulas at large area.

This $F_\alpha[H,\vec r]$ is proportional to $\dal \Phi(\vec r)$. Thus, the quadratic part of our target functional is just a kinetic energy of a free scalar field, but it is the linear term which forces us to use $H(\vec r)$ as an unknown.

It is also worth noting that the energy dissipation $\nu \int_r \oal^2$ is proportional to the same kinetic energy of the scalar field $\phi_1$ on the discontinuity surface. 

In order for $\Phi(\vec r)$ and its gradients to remain finite at the boundary $C$ the new field $H$ should satisfy Dirichlet boundary condition
\begin{equation}
    H(C) = 0
\end{equation}
In order for vorticity to remain finite at the origin we have to have
\begin{equation}
    F_\alpha[H,\vec 0] =0
\end{equation}

Coulomb poles disappeared from this problem, being replaced by weaker, logarithmic singularities (see the next section).

The circulation integral
\begin{equation}
    \Gamma[C] = m Z\int d \theta \left(\Phi\left(R\vec f(\theta)\right) - \Phi(\vec 0)\right)
\end{equation}
with $C:\vec r = L \vec f(\theta)$ being the equation for the contour $C$ in polar coordinates on the plane.

In Appendix F we describe finite element method to solve this variational problem.

\section{Viscosity Anomaly and Scaling Laws}

After rescaling of basic fields the global variable $Z$ only enters in the viscosity term, circulation and the energy balance terms
\begin{eqnarray}\label{InverseZ}
    && G_\alpha =  \frac{\nu}{Z} \d^2\oal + \obe \dbe \val - \vbe \dbe \oal;\\
    && \Gamma(C) = Z \int_S d \sigma_\alpha \oal = Z \oint_C \val d \ral;\\
    && \mathcal{E} = Z^2\int_r \nu \oal^2 = Z f_\alpha Q_\alpha(\vec f)
\end{eqnarray}

The last two global constraints are inserted as delta function in our partition function
\begin{eqnarray}
   &&\mathcal{Z}\left(\Gamma, \mathcal{E}\right)  = \int d \mu(\mathcal{G}) \delta(\Gamma-Z \oint_C \val d \ral)\nonumber\\
   &&\delta\left(\mathcal{E} - Z f_\alpha Q_\alpha(\vec f)\right)\delta\left(\mathcal{E} - Z^2 \int_r \nu \oal^2\right)
\end{eqnarray}
In the linear approximation at small force (zero term vanishes from space symmetry)
\begin{equation}
    Q_\alpha(\vec f) \ra f_\alpha f_\beta Q_{\alpha\beta}
\end{equation}

Let us investigate velocity field in linear vicinity of a \DS{}, with normal distance $z \ra 0$.
We are not going to assume viscous terms to be a small perturbation, just take $z \ra 0$. Nor do we need to assume here that the \DS{} is flat.
The \GBF{} equation for velocity field (with our new normalization)
\begin{eqnarray}\label{GBFV}
   && 0 = \frac{\nu}{Z} \d^2 \val   - \vbe\dbe \val - \dal p ;\\
   && \d^2 p = -\dal \vbe\dbe \val;\\
   && p (\vec r \ra \infty) \ra -\ral f_\alpha
\end{eqnarray}
The boundary condition for pressure is irrelevant at the moment as we investigate this equation in the linear vicinity of the \DS{}.

Before we substitute the singular instanton solution into above \GBF{} equation, we need to smear the theta function.
\begin{eqnarray}
    &&\theta_h(z) = \int_{-\infty}^z d z' \delta_h(z'),
\end{eqnarray}                                                          
where  $\delta_h(z)$   is some approximation to the delta function with width $h \ra 0$.     
The shape of smeared delta function will follow from the \NS{} equations.

The \CL{} representation
\begin{eqnarray}
    &&\val = -\phi_2\dal \phi_1 + \dal \tilde \phi_3;\\
    && \tilde \phi_3 = \phi_3 + \phi_1 \phi_2
\end{eqnarray}
allows us to single out the singular terms in local tangent frame, with $z$ being the normal distance to the surface, and $x,y$ the coordinates in a tangent plane.
\begin{eqnarray}
    &&v_i(x,y,z) = -2 \pi n \theta_h(z) \d_i \Phi(x,y) + \dots ;\\
    && v_z(x,y,z) = z v'_z(x,y) + \dots;\\
    && \d^2 p = -4 \pi n  \d_i \Phi(x,y)\d_i v'_z(x,y)  z \delta_h(z) + \dots;\\
    && p \ra z \delta_h(z) P(x,y) + \dots;\\
    && \d_i^2 P = -4 \pi n  \d_i \Phi(x,y)\d_i v'_z(x,y)
\end{eqnarray}

where $\dots$ stand for a regular parts at $z\ra 0$.

Let us collect most singular terms, proportional to $z\delta_h(z), \delta'_h(z)$ with coefficients depending only of $x,y$:
\begin{eqnarray}
    && 2 \pi n \left( \frac{\nu}{Z } \delta'_h(z) +  v'_z z \delta_h(z) \right)\d_i \Phi\nonumber\\
    &&+ \d_i P z \delta_h(z) =0;\\
    && \d_i^2 P = -4 \pi n  \d_i \Phi\d_i v'_z
\end{eqnarray}
Solving for $P,v'_z $ we find
\begin{eqnarray}
    &&\frac{\nu}{Z}\delta'_h(z) +  v'_z(x,y)z \delta_h(z) =0\\
    && \d_i P(x,y) =0.
\end{eqnarray}
which leads to the Gaussian for normalized distribution $\delta_h(z) =\theta'_h(z)$ and constant solution for $v'_z(x,y)$:
\begin{eqnarray}
    && \delta_h(z) = \frac{1}{ h \sqrt{2\pi}} \EXP{- \frac{z^2}{2 h^2}};\\
    &&v'_z(x,y) = \frac{\nu}{Z h^2};\\
    && P(x,y) =0
\end{eqnarray}

This is viscosity anomaly we were talking about: the singular term $\propto z\delta(z)$ in the Euler equation is balanced by the singular  contribution $\propto \nu \delta'(z)$ from dissipation term. Matching these terms leads to the Gaussian smearing of the delta function.
Now we have to assume some scaling law in the turbulent limit
\begin{eqnarray}
    h \propto \nu^\alpha
\end{eqnarray}
The index $\alpha$ will be determined from  the energy balance equation.

With Gaussian regularization of the delta function we have
\begin{eqnarray}
    &&\int_r \nu \oal^2 \ra \nu \int_r \delta_h(z)^2 \left(2 \pi n \d_i \Phi\right)^2\nonumber\\
    && \ra \Lambda \int_S d^2 r \left(2 \pi n \d_i \Phi\right)^2;\\
    && \Lambda = \frac{\nu}{h}\sqrt{\frac{1}{4  \pi}};\\
    &&\mathcal{E} = Z^2 \Lambda A = Z Q_{\alpha\beta} f_\alpha f_\beta;\\
    && A = \int_S d^2 r \left(2 \pi n \d_i \Phi\right)^2\label{APhi}
\end{eqnarray}
Nonzero solution for $Z$
\begin{eqnarray}
    &&Z = \frac{Q_{\alpha\beta} f_\alpha f_\beta}{\Lambda A} ;\\
    &&\mathcal{E} = \frac{1}{ \Lambda A} \VEV{\left(Q_{\alpha\beta} f_\alpha f_\beta\right)^2}\nonumber\\
    && \propto \frac{h \sigma^2}{\nu}
\end{eqnarray}
From the last relation we finally find the estimate of the random force variance $\sigma$ and pancake width $h$ in the turbulent limit
\begin{eqnarray}
   &&\sigma \sim \nu^{\oh(1-\alpha)};\\
   &&Z \sim \frac{h \sigma}{\nu} \sim \nu^{-\oh(1-\alpha)};\\
   && h \sim \nu^\alpha;\\
   && v'_z(x,y) = \frac{\nu}{Z h^2} \sim \nu^{\frac{3 - 5 \alpha}{2}}
\end{eqnarray}
The self-consistency requires 
\begin{equation}
    \alpha = \frac{3}{5}
\end{equation}
in which case the anomaly contributes to the \NS{} equations in the Turbulent limit.
Restoring powers of $\mathcal{E}, v_z'$ we find:
\begin{eqnarray}
   && h \sim  (v'_z)^{-\nicefrac{9}{10}}\mathcal{E}^{-\nicefrac{1}{5}} \nu^{\nicefrac{3}{5}};\\
   && Z \sim (v'_z)^{\nicefrac{4}{5}}\mathcal{E}^{\nicefrac{2}{5}} \nu^{-\nicefrac{1}{5}};\\
      &&\sigma \sim Q^{-1}(v'_z)^{-\nicefrac{4}{5}}\mathcal{E}^{\nicefrac{3}{5}} \nu^{\nicefrac{1}{5}};
\end{eqnarray}
The dimensional counting seems wrong, but we remember that after our renormalization of \CL{} field, velocity, vorticity and pressure we have following table of dimensions.

\begin{table}[ht]
\tbl{\label{tab:table1}Length-Time dimensions of various variables and parameters.}
{
    \centering
    \begin{tabular}{r||rrrrrrrrrrrr}
     Variable & $\mathcal{E}$ & Z  & $\nu$ & Q &$\omega$ &v &$v'_z$ & h & f & $\sigma$ &A &$\phi$\\ 
     \hline \hline
    Length & 5  & 2 & 2 &9 &-2&-1&-2&1&-3&-6&0&0\\ 
    Time   & -3& -1& -1 &-2&0 &0 &0 &0&0 & 0&0&0\\ 
    \end{tabular}
}
\end{table}

With this table of dimensions the dimensions of above equations all match. In particular, both $Z$ and $\nu$ scale as $L^2/T$ and $\omega$ and $v'_z$ scale as $1/L^2$.

Note that $A$ in above equation \eqref{APhi} is dimensionless as well as \CL{} field. Also note that all renormalized variables and parameters $\oal,\val, f_\alpha, h ,\sigma$ scale as powers of coordinate $r$. Time scale disappeared from our renormalized \GBF{} equations.

Comparing with conventional definitions we see that Reynolds number corresponds to
\begin{equation}
   \mathcal{R} \propto \left(\frac{Z}{\nu}\right)^{\nicefrac{5}{6}}\propto (v'_z)^{\tt} \mathcal{E}^{\ot}\nu^{-1}
\end{equation}
As expected, both the variance and the width go to zero in the turbulent limit. One can estimate the next corrections to the energy balance equation, coming from the $Z$ dependence of vorticity by means of the viscous term in \GBF{} equation. Differentiating $\oal$  by $Z$ and estimating the corrections to $\nu \int_V \oal^2$ we find that these corrections are smaller than the leading terms in the turbulent limit.

\begin{figure}
    \centering
    \includegraphics[width=0.9\textwidth]{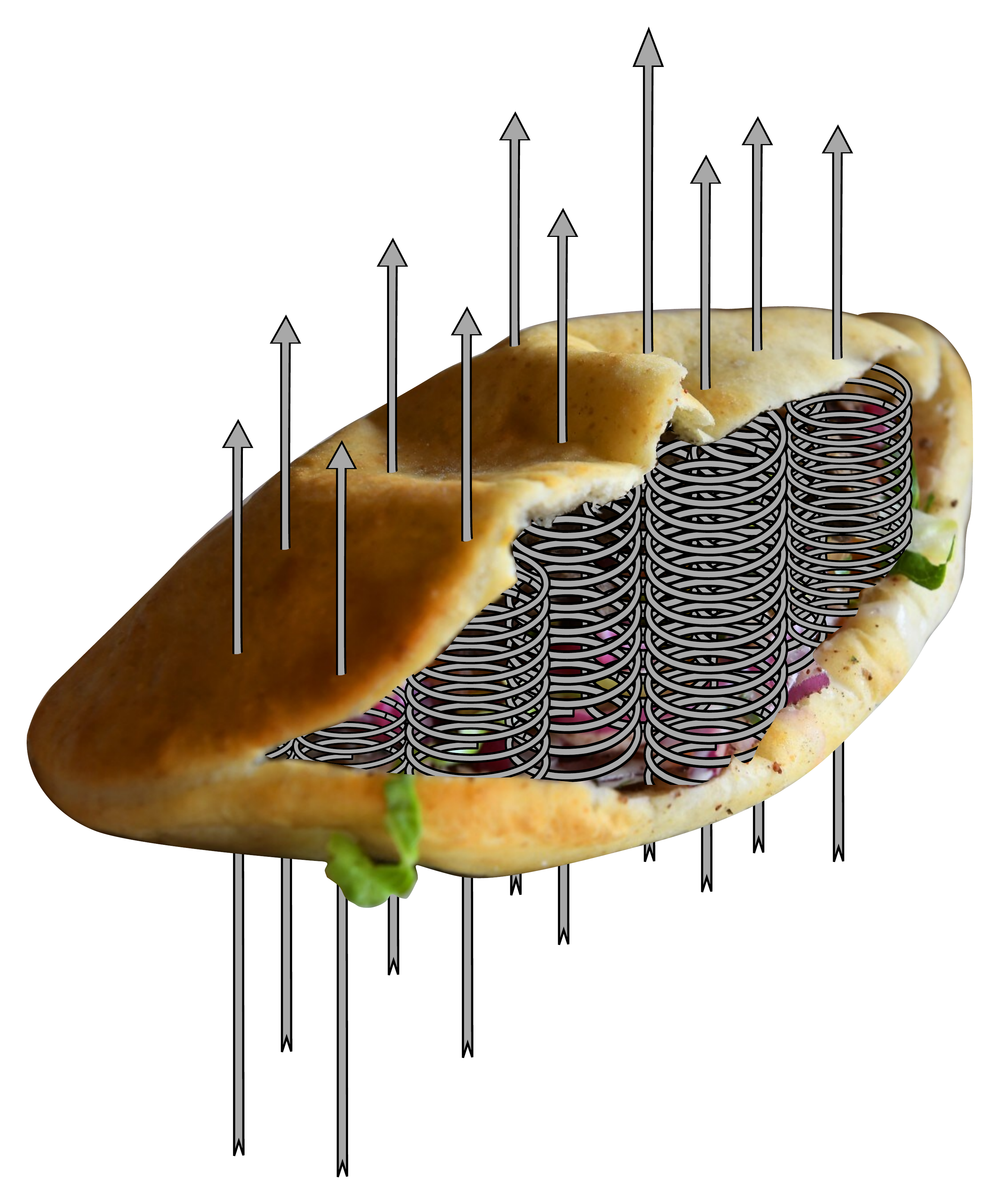}
    \caption{The vortex lines coiling inside the Zeldovich pancake in our Instanton solution.}
    \label{fig::Coil}
\end{figure}

 As for the  Zeldovich pancake, it is filled with coiled vortex lines coming and exiting in the normal direction and making $n$ coils within the thickness $h$ of the pancake (see Fig.\ref{fig::Coil}). 

 The azimuth on our sphere $S_2$ varies as $\varphi = 2\pi n \theta_h(z)$.  In other words this unit vector $\vec S$ makes $n$ rapid rotations around vertical axis, with angle changing as the error function. 
 We study this phenomenon in some detail in Appendix D.
 
\section{Circulation PDF}

In this section we are going to finally derive predictions for the circulation PDF.
\begin{eqnarray}
     &&\Gamma[C] \propto \frac{m}{n}Z\INT{0}{2 \pi} d \theta \int_{D_C} d^2 r \nonumber\\
     &&\frac{H(\vec r)}{\bar H} \left(\frac{1}{|\vec r - L \vec f(\theta)|} - \frac{1}{|\vec r|}\right);\\
     &&\bar H = \frac{\int_{D_C} d^2 r H(\vec r) R(\vec r)}{\int_{D_C} d^2 r  R(\vec r)}
\end{eqnarray}
We remind that the origin is placed at geometric center of the domain $D_C$.

The integral $\int_{D_C} d^2 r H(r) R(\vec r)$ in $\bar H$ is concentrated on finite scales $\vec r \sim 1$ due to decrease of  $R(\vec r)$, so this $\bar H$ scales as $H(\vec 0)$, same as $H(\vec r)$ in the integral in the numerator. 

Collecting scales of the remaining factors we see that $\Gamma[C] = L F[C/L]$ in agreement with the loop equation arguments\cite{M19b}.

Taylor expansion of $\vec Q(\vec f) $ would be justified if, just like in a critical phenomena in statistical physics, the corresponding susceptibility  would grow to infinity to compensate small value of external force. 

This is what happens in a ferromagnet near the Curie point, when infinitesimal external magnetic field is enhanced by large susceptibility, resulting in a spontaneous magnetization.

In our theory this happens because  the pancake thickness $h \propto \nu^{\nicefrac{3}{5}}$ becomes small at together with variance of external force $\sigma \propto \nu^{\nicefrac{1}{5}}$.  The resulting factor $\frac{h}{\nu} \sim \nu^{-\nicefrac{2}{5}}$ enhances the leading term $\left(Q_{\alpha\beta} f_\alpha f_\beta\right)^2 \sim \sigma^2 $ so that the higher terms $ O(\sigma^2)$ of expansion would be negligible. In other words, singularities of the instanton are the origin of the critical phenomena in our theory.

The critical phenomenon, which in our case is the transformation of the Gaussian distribution to an exponential one, happens because of the $\vec Q(\vec f)$ factor  multiplying the Gaussian force in the $Z$ factor in the circulation.

Resulting square of Gaussian variable transforms the Gaussian distribution to  the exponential one.

Also, we observe that the sign of $\Gamma $ is proportional to the sign of the ratio of winding numbers $\frac{m}{n}$. 

Clearly, in addition to solution with winding numbers $m,n$ there are always mirror solutions with $\pm m,\pm n$.

The weight at this solution in our partition function is exactly the same as for the positive $m,n$, so the contributions from these flows must be added. There are also some zero modes related to gauge invariance and conservation of Lagrange multiplier $U_\alpha(\vec r)$ which we integrated out with proper gauge conditions, discussed above and in Appendix C.

This contribution from anti-instantons provides the negative branch of circulation PDF. 

Summing up contribution from both signs we obtain an explicit formula for a Wilson loop 
\begin{eqnarray}
   &&\VEV{\EXP{\i \gamma\Gamma_C}}_{m,n}  = \nonumber\\
   &&\oh \left(W\left(\frac{m}{n}\gamma\right) + W\left(-\frac{m}{n}\gamma\right)\right);\\
   &&W(\gamma) = \frac{1}{\sqrt{\prod_{i=1}^3\left(1 -  \i \gamma \mu_i \Sigma[C]\right)}} 
\end{eqnarray}
where $\mu_i \propto \nu^{\nicefrac{1}{5}}$ are three positive eigenvalues of the matrix (in decreasing order)
\begin{subequations}
\begin{eqnarray}
    &&\mu_{\alpha\beta} = \frac{\sigma Q_{\alpha\beta}}{\Lambda}  \\\label{Sigma}
    &&\Sigma[C] =\INT{0}{2 \pi} d \theta \int_{D_C} d^2 r  \frac{H(\vec r)}{\bar H}\nonumber\\
    &&\left(\frac{1}{|\vec r- L \vec f(\theta)|} - \frac{1}{|\vec r|}\right)
\end{eqnarray}
\end{subequations}
This corresponds to  asymptotic law
\begin{equation}
    P\left(\Gamma\right) \propto \sqrt{\left|\frac{n}{m\Sigma[C]\Gamma}\right|} \EXP{- \left|\frac{n\Gamma}{ m \mu_1 \Sigma[C]}\right|}
\end{equation}


The functional $\Sigma[C]$ is completely universal and calculable in terms of the our universal minimization problem, except for the unknown function $R(\vec r) = v^s_z(\vec r, z)$. Remaining non-universal parameters of the random forces are hidden in the matrix $\hat \mu$. 

This function $v^s_z(\vec r, z)$ is concentrated on the finite sizes near the middle of our domain and falls off as $1/|r|^3$.  Therefore, at large sizes of the loop and the area of the domain $D_C$ this integral can be approximated as
\begin{eqnarray}
&& \int_{D_C} d^2 r v^s_z(\vec r, z) = \mbox{const}\\
  && \bar H \approx H(\vec 0)
\end{eqnarray}
 
The same approximation can be made in the target functional of our minimization problem. After that, the solution for $H(\vec r)$ and $\Sigma[C]$ will be universal.

It is also assumed that the circulation is large compared to the viscosity, and by definition of the WKB approximation we were considering the tails of distribution, at $ |\Gamma| \gg \mu_1 |\Sigma[C]|$.

In that region the  (even) moments $M_p = \VEV{\Gamma^p} $ grow as $\Gamma(p+\oh)$.

Another interesting prediction we have here is a nontrivial dependence of the circulation scale $\Sigma[C]$  from the shape of the loop $C$.

This function can be computed numerically using the variational method we outlined above. In particular, for the rectangle all singular integrals are calculable, so this problem is tractable.

\section{ Topology of Instanton and Circulation PDF}

The quantization of the circulation in a classical problem deserves further attention.

One may wonder what are the physical values of the winding numbers $m,n$. Maybe only the lowest levels are stable, and higher ones must be discarded?

If you consider effective Hamiltonian contribution from this instanton you observe that it does not depend of winding numbers as the solution for $\Phi$ does not depend of $m$ and is inversely proportional to $n$.

Therefore, the circulation  only depends of ratio of winding numbers $ \frac{m}{n}$.
In general case we have to sum over all  $m,n$ with yet unknown weights
\begin{equation}
    \VEV{\EXP{\i \gamma\frac{\Gamma_C}{\sqrt{A_C}}}}  \propto \sum_{m,n \in \mathbb{Z},m,n \neq 0} W\left(\frac{m}{n}\gamma\right) 
\end{equation}

The PDF tail from each term would be
\begin{equation}
    \frac{1}{\sqrt{|\Gamma|\mu_1 \Sigma[C]}} \EXP{- \frac{|n\Gamma|}{|m \mu_1 \Sigma[C]| } } \sqrt{\left|\frac{n}{m}\right|} 
\end{equation}

If we sum over all rational numbers $\frac{m}{n}$ the exponential decay would become power-like contrary to numerical experiments\cite{S19} which strongly support a single exponential. 

So, there is still something we do not understand about our measure on \GBF{} : there are some topological super-selection rules on top of the steadiness of the flow and minimization of effective Hamiltonian.

The conventional helicity integral for our solution is computed and discussed in\cite{M20a} and also in Appendix E of this paper.

Another topological invariant which depends of these winding numbers was suggested in\cite{M20a} where it was argued that it was distinguishing our solution from generic \CL{} field.

\begin{figure}
    \centering
    \includegraphics[width=0.9\textwidth]{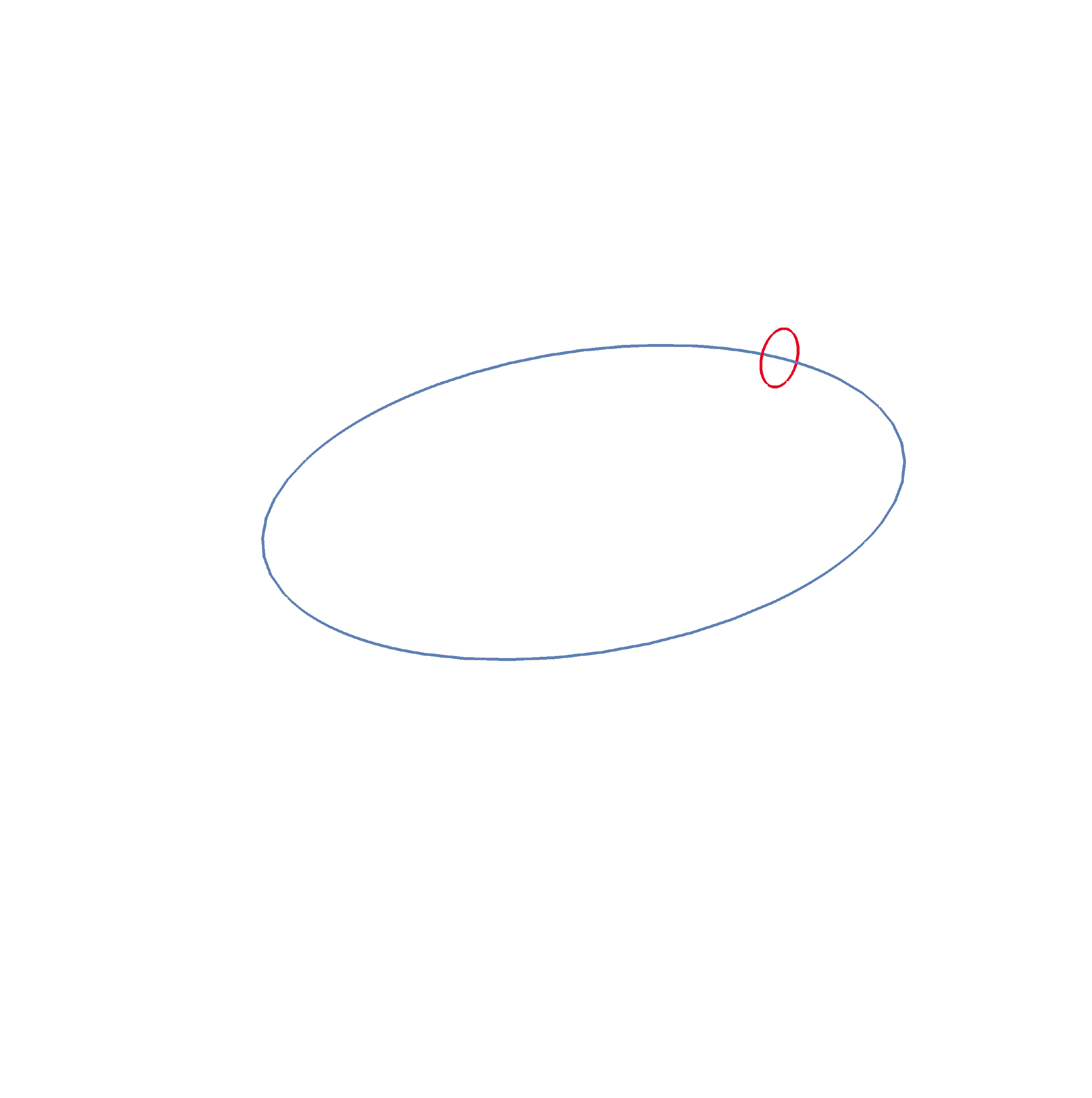}
    \caption{The infinitesimal loop $\delta C$ (red) encircling original loop $C$ (blue).}
    \label{fig::DeltaC}
\end{figure}

Consider the circulation $\Gamma_{\delta C(\alpha)}$ around the infinitesimal loop $\delta C(\alpha)$ which encircles our loop at some point with angular variable $\alpha$ (Fig.\ref{fig::DeltaC}). 
Fig.\ref{fig::DeltaC}
It is straightforward to compute
\begin{equation}
    \Gamma_{\delta C(\alpha)} = \oint_{\delta C(\alpha)} \phi_1 d \phi_2 = 2 \pi n \phi_1
\end{equation}
Clearly, this circulation stays finite  in a limit of shrinking loop $\delta C$ because of singular vorticity at the loop $C$.

Now, integrating this over $d \phi_2 = m d\alpha$ we get our original circulation
\begin{equation}\label{LoopLoop}
   \oint \Gamma_{\delta C(\alpha)} d \phi_2(\alpha) = 2 \pi n \oint \phi_1 d \phi_2 = 2 \pi n \Gamma_C
\end{equation}

\begin{figure}
    \centering
    \includegraphics[width=0.9\textwidth]{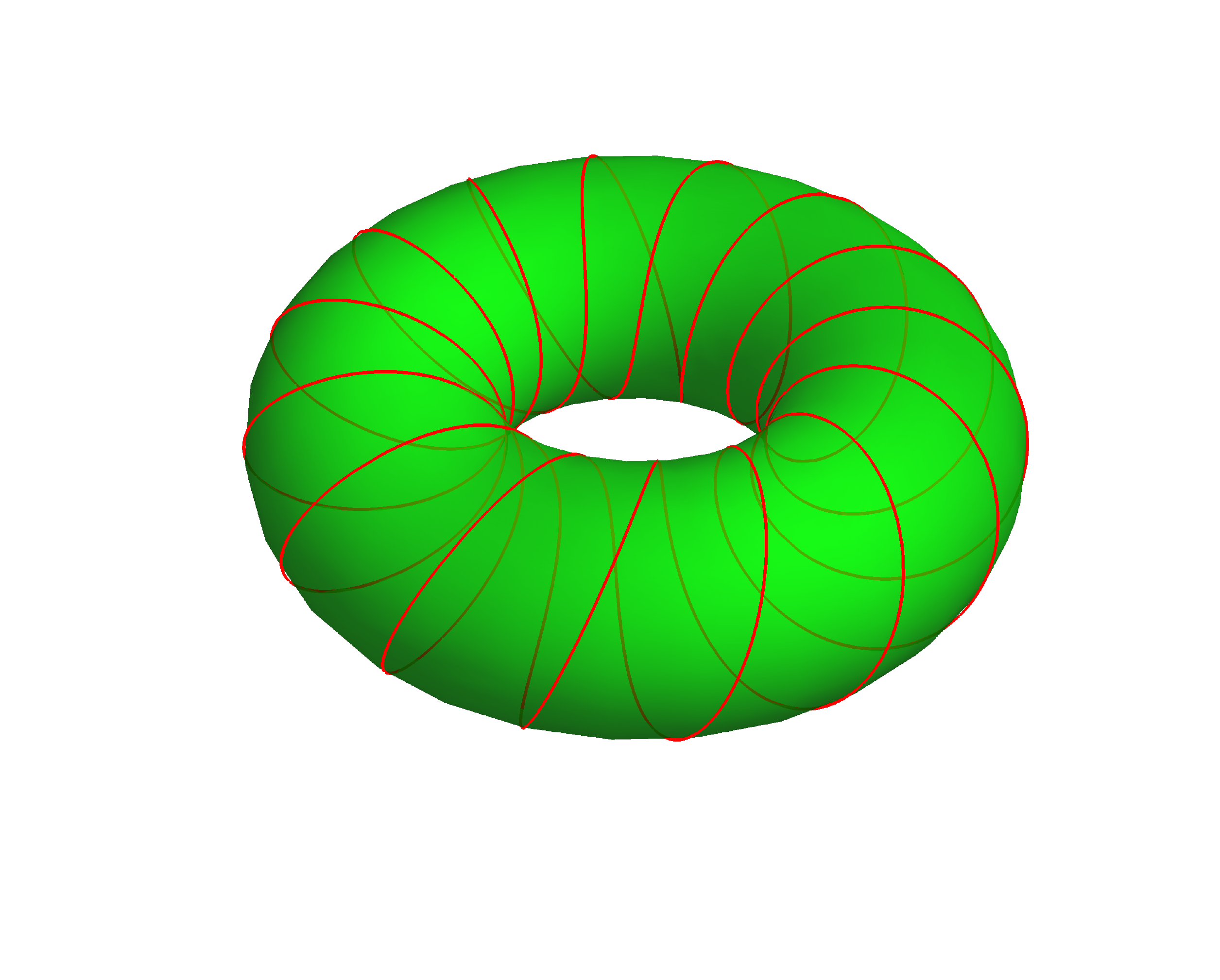}
    \caption{The solid torus mapped into \CL{} space }
    \label{fig::TorusLoop}
\end{figure}

Geometrically, this is a volume of the solid torus in \CL{} space mapped from the tube made by sweeping the infinitesimal disk around our loop (see Fig.\ref{fig::TorusLoop}). 

This volume stays finite in the limit of shrinking tube and equals $2\pi n $ times the velocity circulation $\Gamma_C$ in original space $R_3$.

This circulation by itself is an oriented area inside the loop in \CL{} space, which area is $m$ times the geometric area, as the area is covered $m$ times by the instanton field. 

Let us look at the topology of the mapping from the physical space to the \CL{} space, assuming this space to be $S_2$ as suggested by\cite{KM80, L81}.

 We cut out of $R_3$ the infinitesimal solid torus around our loop -- this remaining space topologically  also represents a solid torus. We cut this solid torus along the \DS{} $S_C$ bounded by $C$, and then glue it back with $2 \pi n$ twist around the polar axis (path inside the solid torus). 
 
 The two sides of the \DS{} are mapped to the spheres $S_2$ which are rotated by $2 \pi n$ around the polar axis. Apparently when we go through the \DS{}  of the viscous thickness $h \sim \nu^{\nicefrac{3}{5}}$ we cover this $S_2$ precisely $n$ times. 

This evolution of $\vec S(x,y,z)$ when $z$ goes from $-h$ to $+h$  describes this rapid rotation around the vertical axis. The tangential vorticity is related to the angular speed of this rotation, which goes to infinity as $1/h$. We discuss this evolution in some detail in Appendix C.

The corresponding vortex lines come from $ z = -\8$, enter the surface at $ z \sim -h$ in the normal direction, then coil $n$ times, then exit at $ z \sim h$ and go to $+\8$ as shown at Fig.\ref{fig::Coil}.

 This is the first cycle. The second one would correspond to the loop around the origin in polar coordinates we used. This contour does not pass through the surface, so it is topologically equivalent to a contractible loop drawn on a surface of this sphere $S_2$.

However, this origin of polar coordinates is not a singularity of our space, this is just a singular system of coordinates.

As we discussed above, near the origin the $\phi_1$ field remains non-singular, with an extra condition $\d_x \phi_1=\d_y \phi_1=0$ at the origin to avoid the $1/|\vec r|$ pole in normal component of vorticity near the surface.

This solid torus with cut surface is topologically equivalent to a 3D ball and our \CL{} field maps this ball onto $S_2$. The winding number $n$ counts the covering of the sphere by this map. 

The second number $m$ would correspond to the periodicity in terms of the angle $\alpha$ in cylindrical coordinates. There is no topological invariant which would protect such a periodic solution.

There is another way to arrive at the same conclusion. Topology of the \CL{} field was analysed in previous work\cite{M20b} (see also Appendix C of this paper) and it was concluded that there is a helicity
\begin{equation}
    H = \int d^3 r \val \oal
\end{equation}
which is characterized by an integer. In Appendix E we compute helicity for our instanton in some general way and we found that it was proportional to the winding number $n$.
\begin{equation}
    H =2\pi n \oint_C \Tilde\phi_3 d \phi_1
\end{equation}
Here $\Tilde\phi_3$ is a third \CL{} field parametrizing velocity
\begin{equation}
    \val = -\phi_2 \dal \phi_1 + \dal \Tilde\phi_3
\end{equation}
This field $\Tilde\phi_3$ is given by the space integral
\begin{equation}
    \Tilde\phi_3 = -\int d^3 r' \frac{\dal \left(\phi_2 \dal \phi_1\right)}{4\pi |\vec r - \vec r'|}
\end{equation}

In virtue of our boundary condition $\d_z \phi_1 =0$ at the singular surface, the delta function in the numerator does not contribute
\begin{equation}
    \d_z \phi_2 \d_z \phi_1 \ra 2 \pi n \delta(z)\d_z \phi_1  =0 
\end{equation}
so that this $\Tilde\phi_3$ is given by non-singular integral over space. It has weak integrated Coulomb singularity at $\vec r' = \vec r$ but no singularity at the \DS{}.
We cannot compute this integral over whole space, but we see that helicity $H$ is proportional to the winding number $n$. 

This supports our argument that $n$ has some topological meaning but $m$ does not.

We therefore restrict ourselves with  solutions with
\begin{equation}
    m = 1
\end{equation}
which have quantized helicity but no fictitious axial singularities.

\section{Discussion. Do we have a theory yet?}

We identified the instanton mechanism of enhancement of infinitesimal random force in Euler equation and demonstrated how this enhancement takes place at small viscosity in Navier-Stokes equation. 

Our view of fluctuating singularity surfaces is dual to conventional picture of fluctuating velocity field with singular correlation in the same way as the weak coupling of the string theory is dual to the strong coupling phase of gauge theories.

An important conclusion from this paper is that turbulence arises spontaneously, with infinitesimal external random forces, as in the ordinary critical phenomena in statistical mechanics. The thickness $h$ of Zeldovich pancakes goes to zero as $\nu^{\nicefrac{3}{5}}$, with tangent components of vorticity approximating a delta function of the normal distance to the surface. The profile is Gaussian with width $h$. 

So, the turbulence is dominated by singular vorticity structures, impossible to describe as interacting waves. The WKB approach, on the other hand, is quite adequate, and it describes most of the PDF of velocity circulation. 

The required random force needed to create the energy flow and asymptotic exponential distribution of circulation, has the variance $\sigma \sim \nu^{\nicefrac{1}{5}}$. This small force is enhanced by large susceptibility $ \sim \nu^{-\nicefrac{1}{5}}$.
This large susceptibility can be traced back to the delta-function singularity of the vorticity field at the minimal surface in the Euler limit of \NS{} equations. 

We presented an explicit solution for the shape of circulation PDF generated by instanton. We claim it is realized in high Reynolds flows for the large loops and large circulations, not as a model, but rather as an exact asymptotic law.

The effective expansion parameter of our weak coupling string theory slowly goes to zero as $\nu^{\nicefrac{1}{5}}$. However, the leading approximation already fits numerical experiments with high accuracy.

We confirmed the dependence $|\Gamma| \propto \sqrt{A_C}$ predicted earlier\cite{M19b} based on the Loop equations. The raw data from\cite{S19} were compared with this prediction. We took the ratio of the moments $M_p = \VEV{\Gamma^p}$ at largest available $p$ and defined the circulation scale as $S = \sqrt{\frac{M_8}{M_6}}$.

Note that the constant term in the effective scaling index $\VEV{|\Gamma|^n} \propto \left(\sqrt{A_C}\right)^{n + \mbox{const}}$ cancels in this ratio of the moments, so we do not determine this constant in our fit.

The DNS for velocity difference scaling laws \eqref{Zeta} suggest that $\zeta(n) \ra \zeta(\infty)$ with rather large limit at $n = \infty$. The circulation would then scale as $ r v$, which would correspond to circulation moments to scale as
\begin{equation}
    M_n \sim \left(\sqrt{A_C}\right)^{n + \zeta(n)}
\end{equation}
So, our scaling law would correspond to finite limit of $\zeta(\infty)$ in agreement with the DNS\cite{ISY20}.

\begin{figure}
    \centering
    \includegraphics[width=0.9\textwidth]{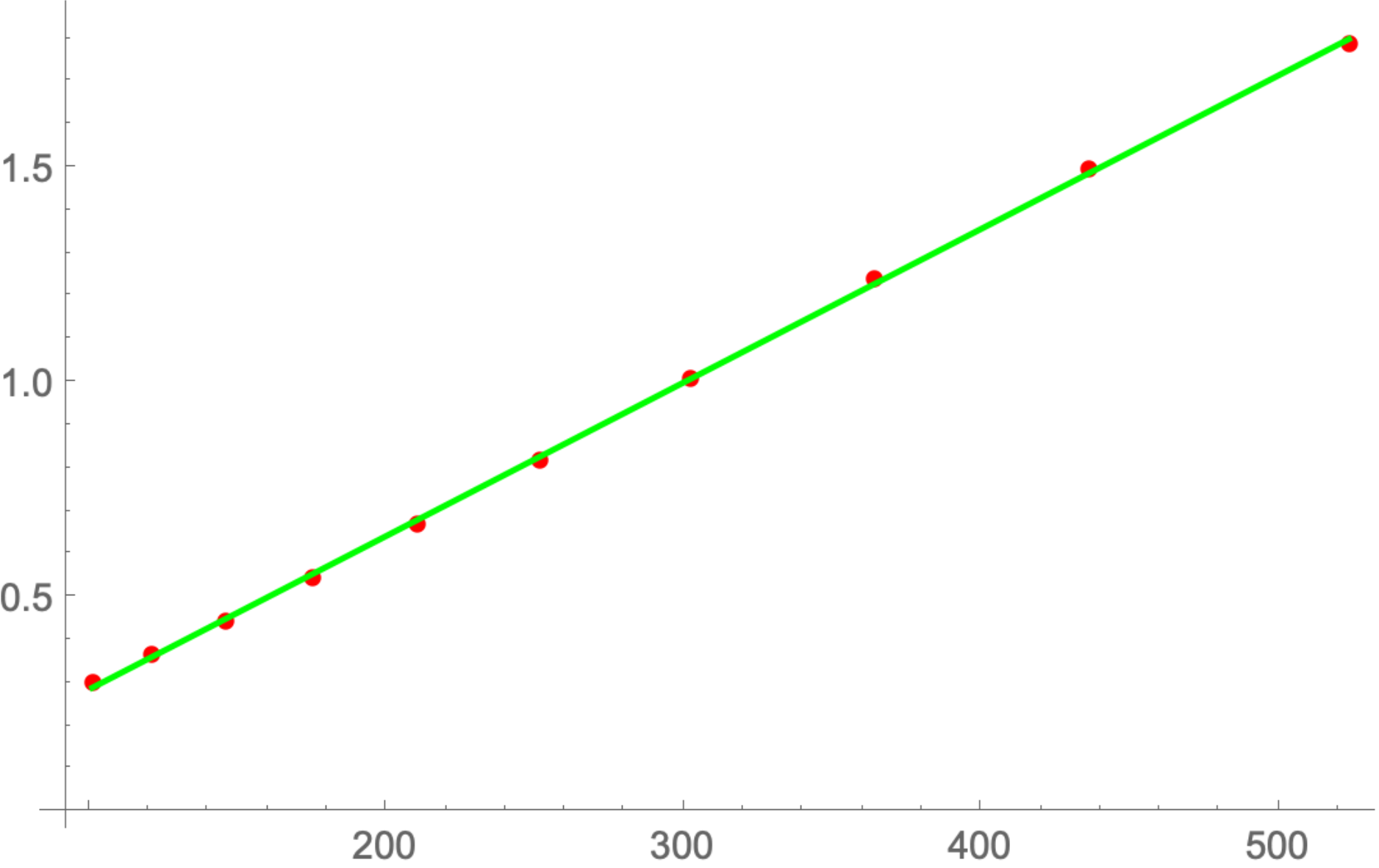}
    \caption{Linear fit of the circulation scale $S = \sqrt{\frac{M_8}{M_6}}$ (with $M_p = \VEV{\Gamma^p}$) as a function of the $R =a/\eta$ for inertial range $ 100 \le R \le 500$. Here $a$ is the side of the square loop $C$ and $\eta$ is a \KO{} scale . The linear fit $S = -0.073404 + 0.00357739 R$ is almost perfect:  adjusted $R^2 = 0.999609$}
    \label{fig::FitSqrtArea}
\end{figure}

We fitted using \Mathematica{} $S(r)$ as a function of the size $r = \frac{a}{\eta}$ of the square loop measured in the Kolmogorov scale $\eta$.
The quality of a linear fit was very high  with adjusted $R^2=0.9996$. 
The linear fit is shown at Fig.\ref{fig::FitSqrtArea}. 
The errors are most likely artifacts of harmonic random forcing at a $8K$ cubic lattice\footnote{This is not to say that some other nonlinear formulas cannot fit this data equally well or maybe even better, for example fitting $\log S$ by $\log R$ would produce very good  linear fit with the slope $1.1$ instead of our $1$. However, this shift of the slope can be imitated by a large intercept $\zeta(\infty)$.
Data fitting cannot derive the physical laws -- it can only verify them against some null hypothesis. This is  especially true in presence of few percent of systematic errors related to finite size effects and harmonic quasi random forcing. We believe that distinguishing between $1.1$ and $1$ is an over-fit in such case.}.

Contrary to some of my early conjectures\cite{M19a}, there is no universality in the area law,
though there is a universal shape of decay of PDF \cite{S19,SK20}, and the singular vorticity at the minimal surface is responsible for that decay. 

The Wilson loop for each winding number is given by a simple algebraic expression
\begin{equation}
   \VEV{\EXP{\i \gamma\Gamma_C}}_n = \frac{1}{\sqrt{\prod_{i=1}^3\left(1 -  \i \frac{\gamma \mu_i \Sigma[C]}{n}\right)}} 
\end{equation}
with $\mu_i$ being a phenomenological parameters but $\Sigma[C]$ in \eqref{Sigma} being calculable in terms of the solution $H(\vec r) $ of universal integral equation, corresponding to minimization of quadratic functional \eqref{Target}.

For the observed rectangular shape  these variation computations can be performed at a supercomputer, so we can compute this function with high accuracy and compare with existing  DNS data.

The PDF is given by sum over positive integer winding numbers $n$ and reduces to well known special function (integral logarithm $\Li_\mu(x)$)
\begin{eqnarray}
    &&P(\Gamma)=\INT{-\infty}{\infty} \frac{d \gamma}{2 \pi}e^{-\i \gamma\Gamma} \VEV{\EXP{\i \gamma \oint_C d \ral \val}} \nonumber\\
    &&\propto \frac{1}{\sqrt{x}}\sum_{n=1}^{\infty}e^{-n x} \sqrt{n} = \frac{1}{\sqrt{x}}\Li_{-\frac{1}{2}}\left(e^{-x}\right);\\
    && x = \frac{|\Gamma|}{ \mu_1 |\Sigma[C]|} ;
\end{eqnarray}

Negative winding numbers are responsible for another branch of the PDF, so that resulting PDF is an even function of circulation at large $|\Gamma|$. 
There are no pre-exponential factors here, as the determinants in the Gaussian functional integral near instanton cancel each other by design. This formula applies at large $x$ which corresponds to the tails of PDF. 

At $\Gamma=0$ there is a singularity, which would require different method to investigate. This corresponds to the tip of the distribution $\Gamma \sim \nu$, where the viscosity cannot be neglected. In the turbulent limit in our theory $\Gamma \sim \nu^{-\nicefrac{1}{5}} \ra \infty$ so that this tip effectively shrinks to zero.

The low moments of circulation are dominated by this tip, where our WKB approximation breaks. For the small enough loop $C$ this will also mean that fluctuations of the singularity surface play the major role. As we suggested in the \eqref{Liouville}, the effective degrees of freedom in that region may be the "Liouville field" coming from fluctuating metric on these random self-avoiding surfaces of singular vorticity.

With effective Liouville action having two parameters $\alpha,Q$ we get a parabolic $\zeta(n)$ which fits existing DNS data up to the maximum at $n\approx 10$.
(see Fig.\ref{fig::Zeta}).

As for the PDF at large loops, we have an exponential law with $1/\sqrt{|\Gamma|}$ factor in front.

\begin{figure}
    \centering
    \includegraphics[width=0.9\textwidth]{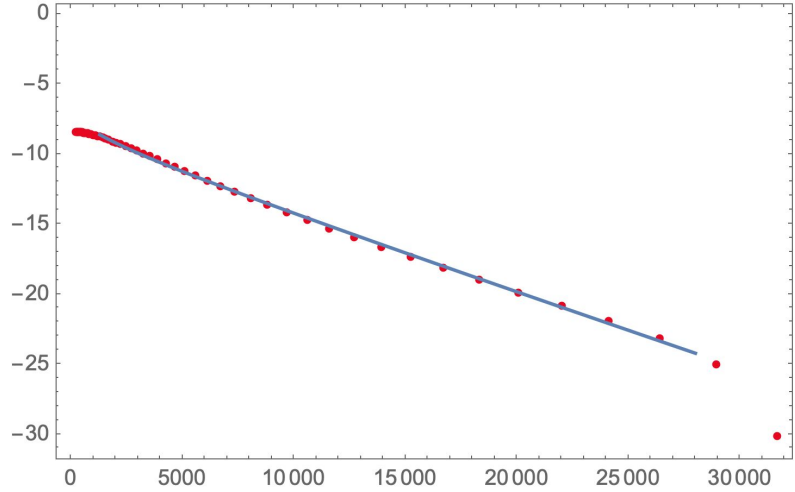}
    \caption{$\log P(x)$ (red dots) together with fitted line $\log P \approx -0.000526724 x-4.3711-0.5 \log (x) \pm 0.116469,\; 1300 <x < 28000$. 
    Here $x = \frac{|\Gamma|}{\nu}$. Last two points have low statistics in DNS and were discarded from fit. Remaining data match the theoretical formula within statistical errors of DNS. Adjusted $R^2 = 0.999929$}
    \label{fig::FitPDF}
\end{figure}

\begin{figure}
    \centering
    \includegraphics[width=0.9\textwidth]{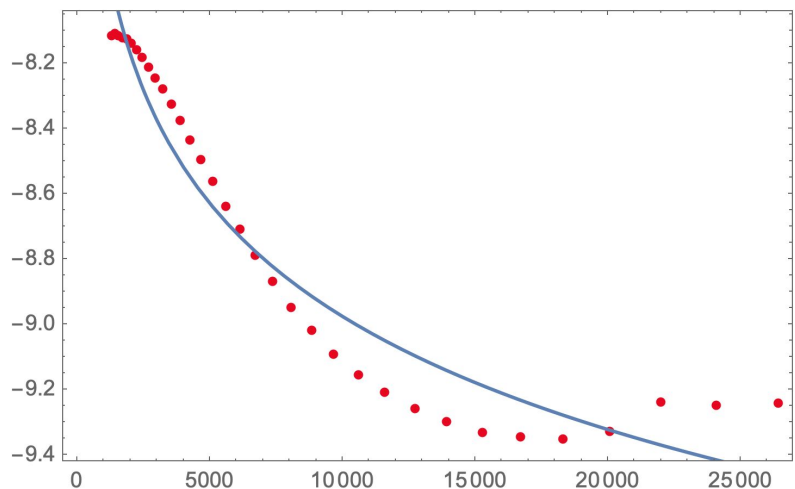}
    \caption{Subtracting the slope. $0.000526724 x +\log P(x)$ (red dots) together with fitted line $ -4.3711-0.5 \log (x),\; 1300 <x < 28000$. 
    Here $x = \frac{|\Gamma|}{\nu}$. We see that the pre-exponential factor $1/\sqrt{|\Gamma|}$ fits the data, though with less accuracy after subtracting the leading term. }
    \label{fig::PreExp}
\end{figure}

\begin{figure}
    \centering
    \includegraphics[width=0.9\textwidth]{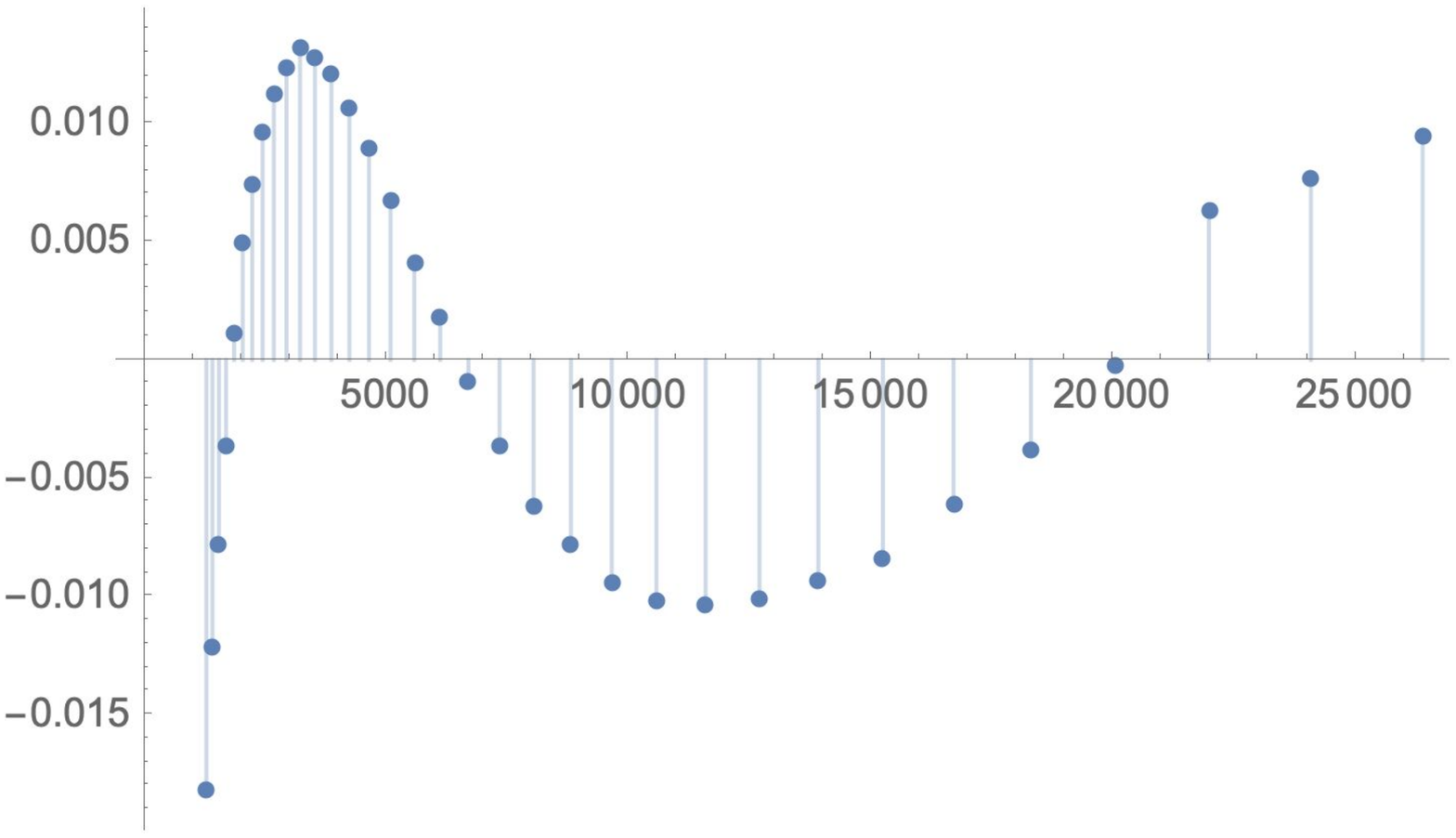}
    \caption{Relative residuals of the log fit of PDF. The harmonic wave behavior suggests that these are artefacts of harmonic random forcing on a $16K^3$ cubic lattice rather than genuine oscillations in infinite isotropic system. Such residuals do not imply contradictions with the theory. }
    \label{fig::Residuals}
\end{figure}

We found that this formula fits the latest data by Kartik Iyer within error bars of DNS with adjusted $R^2 = 0.9999$\footnote{Again, some nonlinear power fit with log/log slope different from $1$ could also fit these data, but as we mentioned above, with systematic errors present we cannot reliably distinguish linear law from power close to $1$.}. See Figs.\ref{fig::FitPDF},\ref{fig::PreExp},\ref{fig::Residuals}.

There is something remarkable with this exponential decay.

With circulation here being the sum of normal components of large number of local vorticities over the minimal surface, it is nontrivial for this circulation to have an exponential distribution, regardless of the local vorticity PDF as long as it has finite variance. 

The Central Limit theorem tells us that unless these local vorticities are all strongly correlated, resulting flux (i.e. circulation) will have a Gaussian distribution. 

The spectacular violation of this Gaussian distribution in the DNS\cite{S19} with seven decades of exponential tails, strongly suggest that there are large spatial structures with correlated vorticity, relevant for these tails.

In this paper, developing and correcting the previous one, we identified these spatial structures as coherent vorticity spread thin over minimal surface. 

We compared the leading term with $n=1$  with this DNS including pre-exponential $1/\sqrt{|\Gamma|}$ factor\cite{M20b}. The detailed comparison was recently performed in\cite{IBS20} with the same positive result.

The sum over integers emerges here by the same mechanism as in Planck's distribution in quantum physics.
There we had to sum over all occupation numbers in Bose statistics.
Here we sum over all winding numbers of the \CL{} field across the minimal surface in physical space.

In Bose statistics the discreteness of quantum numbers is related to the compactness of the domain for the corresponding degree of freedom.

In our case this also follows from compactness of the domain for the \CL{} fields, varying on a sphere $S_2$ . The velocity circulation in physical space becomes the area inside oriented loop on that sphere.

The physical reason why the ambiguous \CL{} fields are acceptable in a real world with single-valued velocity field is the unbroken gauge invariance, or \CL{} confinement. \CL{} fields are ambiguous and unobservable, just like quarks or gluons.

The very tip of this distribution is influenced by the dissipation effects leading to asymmetry of this tip. These effects are given by above viscosity anomaly in the \NS{} equation. 

Also, the random surfaces strongly fluctuate in this region of small circulation. So, some form of Liouville model is adequate here (Fig \ref{fig::SpikyBall}, \ref{fig::Zeta}).

However, in the turbulent limit the scale of circulation $\Gamma_C$ grows, so that this tip with its dissipation shrinks to zero. In extreme turbulent limit the pancakes thickness also shrinks to zero, PDF becomes exponential and we are left with classical instanton on the minimal surface.

It would take very large scale simulations to study these remarkable phenomena because of the slow growth $\nu^{-\nicefrac{1}{5}} \sim \mathcal{R}^{\nicefrac{1}{5}}$ of the circulation scale with Reynolds number. Let us hope that Moore's law (and the leaps of quantum computing\cite{QC20}) will help us simulate these phenomena in near future.

It is possible that similar phenomena exist on the cosmic scale with giant pancakes spanning mega parsecs. We also expect that quantization of our instantons with obvious replacement of the unit vector $\vec S $ by operator of angular momentum with $O(3)$ algebra will lead to some advances of the theory of turbulence in quantum fluids. Our winding numbers $n$ will then become angular quantum numbers.

But the most urgent task is to confirm in DNS the main conjecture that the \GBF{} with random boundary forces is describing the statistics of Turbulence. 

This project is very well defined. Use the same cubic lattice but replace the periodic boundary conditions by our condition for the pressure  $p \ra -\vec f \cdot \vec r$ on a surface of this cube. The steady flow is supposed to be equivalent to the ordinary DNS, which can be verified numerically for the moments of circulation.

So, do we have a theory of turbulence? Not yet IMHO, but we may be getting there. 

Once again I am appealing to young mathematical physicists and string theorists: come and help me! Do not wait for the experts in Turbulence to endorse this theory: they will take forever. 
There is a gauge-string duality in play here, and you know it better.
You would understand it and you can develop it into a Theory of Turbulence.

\section*{Acknowledgments}

I am grateful to Nikita Nekrasov for helping me understand the topology of \CL{} field as well as the properties of the \DS{}.

Sasha Polyakov read the early draft of this paper and we had a productive discussion, helping me understand the meaning of my distribution. 

Useful discussions with Grisha Falkovich, Eugene Kuznetzov, Eugene Levich, Thomas Spenser, Samson Shatashvili and Victor Yakhot helped me understand better the physics and mathematics of this theory.

I also benefited from discussions with Kartik Iyer and Katepalli Sreenivasan regarding numerical simulations. They provided the numerical data used for comparison here. This theory perfectly matches their numerical experiments.

This work is supported by a Simons Foundation award ID $686282$ at NYU. 
\appendix
\section{Finite Dimensional Stationary Distribution}
 Let us study our distribution for a simple example of $N$ dimensional particle moving in phase space $\vec \phi$ with Hamiltonian:
\begin{eqnarray}
    &&\vec \phi = (p_i,q_i)\\
    &&H(\vec \phi) = \frac{\vec p^2}{2} + U(\vec q)
\end{eqnarray}
Let us consider some vector functions $\vec \omega(\vec \phi)$ in phase space which we would like to be stationary so we impose constraints
\begin{equation}
    \vec G = \d_t \vec \omega =0
\end{equation}
 
The steady state equations would be simply :
\begin{subequations}
\begin{eqnarray}
    &&\d_t \vec \phi = (-U_i,p_i);\\
    &&G_\alpha = \pbyp{\oal}{\phi_a}\d_t \phi_a \\
    &&\PF{\PBR{G_\alpha,G_\beta}}  = \sqrt{\det \hat g}\\
    && \hat g_{a b} = \pbyp{G_\alpha}{\phi_a}\pbyp{G_\alpha}{\phi_b}
\end{eqnarray}
\end{subequations}

with $U_i = \d_i U , U_{i j} = \d_i \d_j U$ etc.
Note that the Jacobian $\det U_{i j}$ is not always positive in this Hamiltonian system, but our pfaffian is positive.

We assume now, that just as in case of continuous \GBF{} equations, there are more constraints $ \oal, \alpha =1,\dots M$ than dimension $2 N$ of our phase space, but there are only $2N$ independent constraints because some of these $G_\alpha$ are linearly related.

Let us consider linear vicinity of the stationary point $\phi^*$ solving $\d_t \vec \phi(\vec \phi^*) =0$ and represent the $M$ dimensional delta function as a Fourier integral
\begin{equation}
    \delta(\vec G(\vec \phi))= \int d^M u \EXP{\i \vec u \vec G(\vec \phi)}
\end{equation}
By definition $G(\vec \phi^*) =0$, so we can expand near this stationary point and we get ( with $ \vec \chi =\vec \phi-\vec \phi^* $)
\begin{equation}
    \int d^M u \EXP{\i  u_\alpha \pbyp{G_\alpha}{\phi_a} \chi_a}
\end{equation}

Now we perform singular value decomposition\cite{SVD} of the rectangular matrix $\pbyp{G_\alpha}{\phi_a}$ (which is an pair of orthogonal transformations in left and right spaces preserving volume elements)
\begin{subequations}
\begin{eqnarray}
    &&\vec u = \sum_i \tilde u^i \vec U^i;\\
    &&\vec \chi = \sum_i \tilde \chi^i \vec V^i;\\
    && \det \hat U = \det \hat V = 1,\\
    && u_\alpha \pbyp{G_\alpha}{\phi_a} \chi_a = \sum_i \tilde u^i \lambda_i \tilde \chi^i;
\end{eqnarray}
\end{subequations}

and we are left with integrals over components $\tilde u^i$ with finite eigenvalues $\lambda_i$ which lead to desired result
\begin{eqnarray}
  &&\int' d^M u \EXP{\i  u_\alpha \pbyp{G_\alpha}{\phi_a} \chi_a}=\nonumber \\
  &&\int' d^M \tilde u \EXP{\i  \sum_i \tilde u^i \lambda_i \tilde \chi^i} \propto \nonumber \\
  &&\frac{\delta^{2 N}(\vec \chi)}{\prod' |\lambda_i|} = \frac{\delta^{2 N}(\vec \chi)}{\sqrt{\det \hat g}}
\end{eqnarray}

The integrals over the zero modes produce infinities and has to be eliminated by our prescription with the Pfaffian.

Following our prescription in this case would lead to the distribution:
\begin{equation}
    P(\vec \phi) = \sqrt{\det \hat g}\delta(\vec G) \propto \sum_{\vec \phi^*:\d_t \vec \phi(\vec \phi^*)=0}\delta(\vec \phi-\vec \phi^*)
\end{equation}
which corresponds to the sum over all equilibrium states.
Each such state $\vec \phi^* = (\vec 0, \vec r)$ corresponds to a particle sitting at the local extremum $\vec r$ of the potential well with zero momentum, with net zero force acting at it.

Note that we count each such equilibrium state (stable or not!) with equal weight, which we normalize to $1$.

In case there is some extra invariance of observables $\vec \omega$ with respect to transformation of original phase space coordinates $\vec \phi$, there will be some zero modes in the metric tensor $\hat g$. 

Integrating over these zero modes (gauge orbits) is not Gaussian, and has to be fixed by some gauge conditions with proper Faddeev-Popov Jacobian, which we do not consider here, as this is a well known procedure.

As for the time independence of the measure, this degeneracy does not affect it: each of these degenerate points does not move in Hamiltonian dynamics, regardless the fact that observables related to these points have the same values.

One could argue that prescription without absolute value of the Jacobian also has mathematical meaning, representing a topological invariant.  In this case the meta-stable states with negative Jacobian will enter with negative sign. 

For example, in one-dimensional case 
\begin{equation}
    \int d x U''(x) \delta\left(U'(x)\right)
\end{equation}
one can start with an oscillator potential $U(x)= \oh x^2$ with only one minimum at the origin and add cubic and quartic terms, leading to the double-well potential with one maximum and two minima. Our pfaffian $|U''(x)|$ would count $1 + 1 + 1 =3$ states in such a system, but the topological prescription would still have $1 -1 + 1 =1$, same as for an initial oscillator. 

The time-independence of this measure is obvious, as the stationary points by definition do not move with time 
\begin{equation}
    \d_t \vec \phi(\vec \phi^*) =0
\end{equation}

Our canonical ensemble would be:

\begin{eqnarray}
    &&\int d^{2 N} \phi \EXP{-\lambda H_{eff}\left(\vec \omega(\vec \phi)\right)} P(\vec \phi) \propto\nonumber\\
    &&\sum_{\vec \phi^*:\d_t \vec \phi(\vec \phi^*)=0}\EXP{-\lambda H_{eff}\left(\vec \omega\left(\vec \phi^*\right)\right)}
\end{eqnarray}

This is an example of so called "trivial" conservation laws, present in every Hamiltonian dynamics: place the system in its mechanical equilibrium, give it zero velocities and it will stay there.

Except in case there are many (or a continuous manifold) of these stationary states, our distribution gives equal weight to each of them. It is implied that the invisible forces from thermostat kick the system from one stationary state to another one, eventually leading to this uniform distribution over stationary states.

In the context of \GBF{} this space of stationary points is not so trivial, in fact, as we shall see it is rich enough to describe the critical phenomena in turbulent flow.

Even in this elementary example we see a complication. Consider axial symmetric potential of sombrero hat.
\begin{equation}
    U = \oh\left( \vec q^2-1\right)^2
\end{equation}
There is a maximum at the origin and degenerate minimum: a sphere $\vec q^2 =1$. 
We get zero determinant at $N>1$ at the minimum because of the zero modes corresponding to rotations of this minimal sphere.

This is clearly not what we need: to reject the maximum and keep the minimum even when it is degenerate.

Say, in one-dimensional example we need only $2$ of $3$ states, rather than the pfaffian counting $3$ or topological counting $1$.

To reject the maximum we need to demand that the whole matrix of second derivatives is positive definite.

To remove the fictitious zero weight, let us add a linear force, which will act as gauge fixing
\begin{equation}
     U = \oh\left( \vec q^2-1\right)^2 - \vec f. \vec q
\end{equation}
Now, at arbitrary $f$ there will be only one stable minimum and we shall pick it, and we can tend $\vec f \ra 0$.

\section{Saddle Point Integral for Energy Surface Constraint}
We reproduce here the transformation from micro-canonical (delta function for conserved global quantity) to the canonical (exponential of this quantity times Lagrange multiplier) using conserved energy $H(p,q)$ as an example.

Let us consider the micro-canonical distribution for some large system consisting of the subsystem $H_1 = H[p_1,q_1]$ with phase space volume $d \Gamma_1 = d \Gamma[p_1,q_1]$ and a thermostat $H_2 = H[p_2,q_2]$ with phase space volume $d \Gamma_2= d \Gamma[p_2,q_2]$:
\begin{equation}\label{Z}
    Z = \int_C \frac{d \lambda}{2 \pi \i} \EXP{\lambda E} \int d \Gamma_1 \EXP{- \lambda H_1} \int d \Gamma_2 \EXP{- \lambda H_2} 
\end{equation}
The integration contour $C$ here goes along the imaginary axis, providing thus the Fourier representation of the delta function which constraints the distribution to the energy surface $H_1 + H_2 = E$.

In case of the ordinary statistical mechanics $H[p,q]$ is the Hamiltonian and $d \Gamma[p,q] = \prod d p d q$  is the linear phase space volume. Resulting distribution would be an ordinary Gibbs distribution $ \EXP{- \beta H_1}$.

However, the mathematical mechanism behind this transformation from the delta function to the exponent is fairly general. It applies to arbitrary (maybe nonlinear) measure $d \Gamma$ and arbitrary (maybe non-positive) Hamiltonian as long as it is bounded from below in the infinite phase space.

In case of Turbulence we are applying this transformation to the system where $[p,q]$ stand for \CL{} variables parametrizing vorticity $\vec \omega = \nabla p \times \nabla q$ and $H[p,q]$  is some positive conserved quantity such as the volume inside the \CL{} discontinuity surface.

The measure $d \Gamma[p,q] $ in case of proposed Field Theory of Turbulence is restricted to so called Generalized Beltrami Flow. Explicit form of this measure is not relevant.

Finally, the energy $E_0 =V$ is the total volume of fluid in case of turbulence.

After all these comments we can proceed with computation, and it goes the same way in both cases: Gibbs and Turbulence.

Namely, we are looking for a saddle point in the one-dimensional integral over $\lambda$.
\begin{align}
    &Z = \int d \Gamma_1 \Omega[p_1,q_1];\\
    &\Omega[p_1,q_1] = \int_C \frac{d \lambda}{2 \pi \i} \EXP{\lambda E- \lambda H_1 + S(\lambda)} \\
    &\EXP{S(\lambda)} = \int d \Gamma_2 \EXP{- \lambda H_2} 
\end{align}

The saddle point equation
\begin{align}
    &\Omega[p_1,q_1] \propto \frac{1}{\sqrt{S''(\lambda)}} \EXP{\lambda E- \lambda H_1 + S(\lambda)};\\
    &S'(\lambda) + E - H_1 =0
\end{align}

Now, assuming that $H_1 \ll H2$ (there is an infinite thermostat $H_2$ and finite subsystem $H_1$ under study) we can approximate $\lambda$ as solution of universal equation (independent of $p_1,q_1$)
\begin{equation}
    S'(\lambda_0) + E =0;
\end{equation}
After that, up to universal factors
\begin{equation}
    \Omega[p_1,q_1] \propto \EXP{- \lambda_0 H_1}
\end{equation}

Now, this $\lambda_0$ in case of thermodynamics is given by inverse temperature  $\lambda_0 = \beta$. In our case it is some parameter characterizing the thermostat $H_2$. By varying the energy pumping to the thermostat we can vary this parameter in the same way as we vary the temperature in the thermodynamics.

As it is evident from this computation, this saddle point, if it exists, can only be at real positive $\lambda_0$, as the integral $\int d \Gamma_2 \EXP{- \lambda H_2} $ converges only in the right semi-plane.

Now, the saddle point equation can be also rewritten as
\begin{equation}
    E(\lambda_0) = \VEV{H_2} = \frac{\int d \Gamma_2 H_2\EXP{- \lambda_0 H_2}}{\int d \Gamma_2 \EXP{- \lambda_0 H_2}}
\end{equation}

Note that $E(\lambda_0)$ is a monotonously decreasing function as
\begin{equation}
    E'(\lambda_0) = - \VEV{\left(H_2 -\VEV{H_2}\right)^2} < 0;
\end{equation}

This also means that the entropy $S(\lambda)$ is convex function
\begin{equation}
    S''(\lambda) = -E'(\lambda) >0
\end{equation}

As a consequence, the factor of $\i$ in $\frac{1}{\sqrt{-S''(\lambda)}} = \frac{\i}{\sqrt{S''(\lambda)}}$ which arises from the Gaussian integration around the saddle point , cancels the factor of $\i$ in denominator of original integral. 

To be more precise, when we move the integration contour $C$ to the saddle point, we have to direct it along the steepest descent path. In our case this path goes in imaginary direction, as the second derivative $S''(\lambda)$ is positive.  Thus, we have 
\begin{align}
    &\Omega[p_1,q_1] = \INT{-\infty}{\infty} \frac{d z}{2 \pi} \EXP{ (\lambda_0 + \i z) (E - H_1) + S(\lambda_0 + \i z)} \\
    &\ra Z_0 \EXP{-\lambda_0 H_1};\\
    &Z_0 =\sqrt{\frac{2 \pi}{S''(\lambda_0)}}\EXP{\lambda_0 E + S(\lambda_0)}
\end{align}
 
Therefore the Gibbs weight $\Omega[p_1,q_1]$ is real and positive as it should be.

Let us now study the important issue of existence and uniqueness of this saddle point.

In case of the thermodynamics as well as for the volume bounded by closed \DS{} in turbulence $H_2$ is positive definite, and by varying $\lambda_0$ along positive axis we go from the region of high energies (small $\lambda$) to the region of low energies (high $\lambda$). So, the expectation value monotonously varies from zero to infinity and at some point it crosses the level $E$ (only once).

This concludes our proof.

\section{Spherical Gauge}
 
 The symmetric metric tensor $g_{i j}$ in 2 dimensions has three independent components: two diagonal values $g_{1 1}, g_{2 2}$ and one off-diagonal value $g_{1 2} = g_{2 1}$. 
 
 We take stereographic coordinates $z = z_1 + \i z_2 = \tan\frac{\theta}{2}  e^{\i \varphi}$
 \begin{subequations}
\begin{eqnarray}
    && g_{i j} = \delta_{i j} \rho;\\
   &&\rho = \frac{1}{\left(1 + |z|^2\right)^2};\\
   && z_a = \frac{S_a}{1 + S_3};\\
   && S_a = \frac{2 z_a}{1 + |z|^2};\\
   && S_3 = \frac{1-|z|^2}{1 + |z|^2};\\
   && d^2 S = d z_1  d z_2 \rho
\end{eqnarray}
\end{subequations}

The $O(3)$ rotation in these coordinates reads (with $I,J,K = 1,2,3, a,b,c...= 1,2$)
\begin{subequations}
 \begin{eqnarray}
 &&\delta S_I = e_{I J K} S_J \alpha_k; \\
 && \delta z_a = \alpha_3 e_{a b} z_b -\oh(1-|z|^2)\tilde \alpha_a  - \tilde \alpha_b z_b z_a;\\
 && \tilde \alpha_b  = e_{b c} \alpha_c;
  \end{eqnarray}
\end{subequations}
The $O(3)$ transformation of the metric tensor involves the matrix $R_{i j} = \d_j \delta z_i$
\begin{eqnarray}
    &&R_{i j} =  \alpha_3 e_{i j} + \tilde \alpha_i z_j - \tilde \alpha_j z_i - \delta_{i j}\tilde \alpha z ;\\
    && \delta_{O(3)} g_{i j} = R_{a i} g_{a j} +  R_{a j} g_{a i}
\end{eqnarray}

Computing the variation $\delta_{O(3)} g_{i j}$ of the conformal metric $g^c_{i j} = \rho \delta_{i j}$ we find
\begin{subequations}
\begin{eqnarray}
    &&\delta_{O(3)} g^c_{1 2} = 0;\\ 
    &&\delta_{O(3)} g^c_{1 1} = -2 \rho \left(z_1 \alpha_2 - z_2 \alpha_1 \right);\\ 
     &&\delta_{O(3)} g^c_{2 2} = -2 \rho \left(z_1 \alpha_2 - z_2 \alpha_1 \right);
\end{eqnarray}
\end{subequations}
 The gauge transformation of conformal metric produces
 \begin{subequations}
 \begin{eqnarray}
    && \delta_{gauge} g^c_{1 2} = (h_{2 2} - h_{1 1})\rho;\\
    && \delta_{gauge} g^c_{1 1} = 2 h_{1 2} \rho;\\
    && \delta_{gauge} g^c_{2 2} = -2 h_{1 2} \rho;\\
    && h_{i j} = \d_i\d_j h(z_1, z_2)
 \end{eqnarray}
\end{subequations}

Now, we do not want to break rotational invariance of the spherical metric. 
This means that  any $h(z)$ satisfying the equations
\begin{eqnarray}
    && h_{2 2} - h_{1 1} = 0;\\
    && h_{1 2}  = -(z_1 \alpha_2 - z_2 \alpha_1) ;\\
    && - h_{1 2}  = -( z_1 \alpha_2 - z_2 \alpha_1) 
\end{eqnarray}
with some finite constant $\alpha_1, \alpha_2$ should not be restricted by our gauge conditions. Adding the last two equations we immediately see that there are no such gauge functions which could imitate the $O(3)$ rotations.

The independent conditions $h_{1 2} =0, h_{1 1} = h_{2 2}$ combine into one complex equation
\begin{equation}
    \hat L h = \rho \frac{\d^2}{\d \bar z^2}h =0
\end{equation}
These gauge conditions leave out arbitrary linear function $h = A + B_i z_i$, corresponding to constant shifts of \CL field. These constant shifts can be fixed by placing the origin at the South Pole which we did.

For remaining nontrivial \SYM{} we have the gauge fixing Gaussian integral 
\begin{subequations}
\begin{eqnarray}
   &&\int D \lambda D \mu D h \EXP{\i \int d^2 S \lambda \hat L_1 h+ \mu \hat L_2 h} ;\\
   && \hat L_1 h = \rho \Re \hat L h;\\
   && \hat L_2 h = \rho \Im \hat L h;
\end{eqnarray}
\end{subequations}
The regularized determinant $\det{\hat L}$ is a universal number, which does not depend on our dynamical variables.

This operator being non-Hermitean, we are not sure how to regularize and compute this determinant, but this is immaterial, as it does not depend on dynamic variables and thus drops from the measure.
 
\section{Winding numbers}

Our singular variables where $\phi_2$ is related to the angular variable in cylindrical coordinates and has $2 \pi n$ discontinuity on a \DS{} raises obvious questions: maybe this is all an artefact of singular coordinates? What happens in a regular gauge where the \CL{} field is continuous?

Let us study the \CL{} field as a point on $S_2$, using the KM parametrization \eqref{S2param}.
The unit vector $\vec S \in S_2$ will have components
\begin{subequations}
\begin{eqnarray}
    && S_3 = 1-\phi_1;\\
   && S_1 + \i S_2 = \sqrt{1-S_3^2} e^{\i \phi_2};\\
   && \oal \propto e_{\alpha\beta\gamma} e_{i j k} S_i \dbe S_j \dga S_k
\end{eqnarray}
\end{subequations}
 As the $2 \pi n$ discontinuities of $\phi_2 $ now "disappeared" in phase factor, how do we get our singular vorticity in this gauge?
 
 Let us resolve this paradox in a physicist's way. These discontinuities are, in fact, the approximation to the peaks of vorticity in Zeldovich pancakes. The \CL{} fields are not discontinuous with finite viscosity, they are rather changing in a thin lawyer of the thickness $h \sim \nu^{\nicefrac{3}{5}}$, imitating step function in a phase discontinuity. 
 \begin{subequations}
\begin{eqnarray}
     &&\phi_2 \approx m \alpha + 2 \pi n\theta_h\left(z\right) + O(z^2);\\
     && \theta_h(z) = \frac{1 +\erf\left(\frac{z}{h\sqrt 2}\right)}{2}
\end{eqnarray}
\end{subequations}
 
 The complex field $\Psi(x,y,z) = S_1 + \i S_2$ now has some rapid changes in the region $ |z| \sim h$ in normal direction to the \DS{}.
 Specifically, we have
 \begin{equation}
     \pbyp{\Psi}{z} = 2 \pi \i n \Psi \theta'_h(z)  + \mbox{reg terms}
 \end{equation}

The vorticity will have singular tangential  components (with all factors $\sqrt{1-S_3^2}$ cancel thanks to \SYM{} invariance of this representation)
\begin{equation}
    \oal \propto 2 \pi  n  e_{\alpha \beta 3} \dbe S_3  \theta'_h(z)  \overset{h \ra 0}{\longrightarrow} \pi  n e_{\alpha \beta 3} \dbe S_3 \delta(z)
\end{equation}

This smearing of a delta function exposed an interesting phenomenon. The two sides of the \DS{} are mapped to the spheres $S_2$ which are rotated by $2 \pi n$ around the $z$ axis. Apparently when we go through the \DS{}  we cover this $S_2$ precisely $n$ times. 

This evolution of $\Psi(x,y,z)$ when $z$ goes from $-h$ to $+h$  describes this rapid rotation of $\vec S(x,y,z)$ around the vertical axis. The tangential vorticity is related to the angular speed of this rotation, which goes to infinity as $1/h$.

The corresponding vortex lines come from $ z = -\8$, enter the surface at $ z \sim -h$ in the normal direction, then coil $n$ times, then exit at $ z \sim h$ and go to $+\8$ as shown at Fig.\ref{fig::Coil}.

There is still a potential singularity in this representation, namely at the axis of cylindrical coordinates, where the plane coordinates $ x + i y \ra 0$. Representing 
\begin{equation}
    e^{\i \alpha} = \frac{x + \i y}{\sqrt{x^2 + y^2}}
\end{equation}
and combining the square roots we have
\begin{equation}
    S_1 + \i S_2 = \sqrt{\frac{1 - S_3^2}{\left(x^2 + y^2\right)^m}} \left( x + \i y\right)^m \EXP{2\pi\i \theta_h(z) + \dots}
\end{equation}

This expression will have no singularities in coordinate space provided near this axis $x,y=0$
\begin{equation}
    S_3^2 \ra 1- \left(x^2 + y^2\right)^m f^2(x,y,z)
\end{equation}

In other words the axis of the cylindrical coordinates maps into one of the poles of the sphere $S_2$. In general case of the non-planar \DS{} this axial axis would be some path intersecting the surface in the normal direction and going to infinity.

So, we view our physical space as the solid torus ($R_3$ with infinitesimal tube around $C$ cut out of it). This solid torus is cut across this \DS{} and glued back with $2 \pi n $  twist around the angle $\alpha$ around the axial origin (path in this solid torus crossing the \DS{}).

One could present a manifestly regular parametrization of the sphere, adequate to our instanton solution, in terms of the stereographic coordinates
\begin{subequations}
\begin{eqnarray}
    && S_3 = \frac{1 - |u|^2 |w|^2}{1 + |u|^2 |w|^2};\\
   && S_1 + \i S_2 = \frac{2 u w}{1 + |u|^2 |w|^2};\\
   && u = (x + \i y)^m;\\
   && \arg w = \phi_2 - m \alpha;
\end{eqnarray}
\end{subequations}

The complex field $w(x,y,z)$, parametrizing the point $\vec S \in S_2$ is single-valued, and does not have any singularity in $x y z$ space, except that its phase rapidly rotates $n$ times around when the surface $S$ is crossed.

This solid torus with the cut is now topologically equivalent to a ball (inside of $S_2$ sphere).
This ball is mapped on a stereographic sphere $S_2$  with its pole corresponding to that axial path. 
The field does not have a singularity st this path.

The winding number $n$ is counting covering of the  sphere $S_2$ in this map from the ball and the number $m$  would count periodicity or the \CL{} field with respect to the cylindrical axis rotation. Generic case would be $m =1$, in which case no adjustment of parameters would be needed to cancel derivatives of $S_1 + \i S_2$ at the cylindrical axis $x=y=0$.

\section{Helicity}

Let us now look at the helicity integral
\begin{equation}
    H = \int_{R_3 \setminus S_{\mbox{min}}}  d^3 r \vec v \vec \omega
\end{equation}

Note that in conventional form
\begin{equation}
    v_i = \phi_1 \d_i \phi_2 + \d_i \phi_3
\end{equation}
there will be singular terms in velocity $\propto \delta(z)$. However, the Biot-Savart integral \eqref{BSvelocity} demonstrates that these singular terms cancel between $\phi_2$ and $\phi_3$ leaving finite resulting velocity field.

To avoid these fictitious singularity, let us rewrite velocity in an equivalent form
\begin{eqnarray}
    &&v_i = -\phi_2 \d_i \phi_1 + \d_i \Tilde\phi_3\\
    &&\Tilde\phi_3 = \phi_1 \phi_2 + \phi_3
\end{eqnarray}
This  $\Tilde\phi_3$ is single-valued, unlike the $\phi_3$. The discontinuity of the first term is compensated by that of the second one. In can be written as an integral over the whole space
\begin{equation}\label{TildePhi3}
    \Tilde\phi_3(r) = - \dbe \int d^3 r' \frac{\phi_2(r') \dbe \phi_1(r')}{4 \pi |r-r'|}
\end{equation}

Now the singular component $\phi_2$ is not differentiated, so that there are no singularities. 
The  helicity integral could now written as  a map $R_3 \mapsto (\phi_1,\phi_2,\Tilde\phi_3)$
\begin{eqnarray}
    &&H =\int_{R_3 \setminus S_{\mbox{min}} } d^3 r \left(-\phi_2 \d_i \phi_1 + \d_i \Tilde\phi_3\right) e_{i j k } \d_j \phi_1 \d_k \phi_2 \nonumber\\
    &&= \int_{R_3 \setminus S_{\mbox{min}} } d\phi_1 \wedge d\phi_2 \wedge d\Tilde\phi_3 \label{fullHel}
\end{eqnarray}

Here is the most important point. There is a surgery performed in three dimensional \CL{}  space: an incision is made along the surface $\phi\left(S_{\mbox{min}}\right)$ and then it is glued back with $2 \pi n$ twist around the axis of cylindrical coordinates.

Integrating over $\phi_2 $  in \eqref{fullHel}, using discontinuity
\begin{equation}
    \Delta \phi_2\left(S_{\mbox{min}}\right) = 2 \pi n
\end{equation} and then integrating 
\begin{equation}
    \int_{S_{\mbox{min}}} d \Tilde\phi_3 \wedge d \phi_1
\end{equation}
we find a simple formula
\begin{equation}\label{Helicity}
    H =2\pi n \oint_C \Tilde\phi_3 d \phi_1
\end{equation}

One may wonder how can the pseudoscalar invariant like helicity be present in \GBF{}: it is just the time reversal which is  broken by energy flow, but not spacial parity.

The answer is that in virtue of the symmetry of the master equation there is always a \GBF{} with an opposite helicity (negative $n$) and the same probability. We will take both solutions, instanton and anti-instanton into account when using the WKB methods to compute circulation PDF.

One may also wonder how do we get the nontrivial helicity if the velocity is orthogonal to vorticity at the surface where all action is happening. There are  two answers.

Formally, helicity is created just by the discontinuity of the \CL{}{} field by the tangent component of vorticity in the infinitely thin boundary layer. This delta function contributes to the helicity integral.

Another answer is that in the helicity integral over the remaining space $R_3 \setminus S_{\mbox{min}}$, the dot product $\vec v \vec \omega$ is not zero but but rather reduces to a total derivative of the phase field $\phi_2$. After cancellations of all internal terms this integral is proportional to the total phase change from one side of the surface to another, which is $2 \pi n$.

Regardless how we compute helicity we observe that resulting loop integral \eqref{Helicity} involves non-singular field $\Tilde\phi_3$ which depends upon the behavior of the basic \CL{} field $\phi_1, \phi_2$ in the whole remaining space, not just in linear vicinity of the \DS{}.

Our main physical assumption was that vorticity was concentrated in a thin layer surrounding the \DS{}.
There is a singular tangential component $\propto \delta(z)$ and smooth normal component. For the smooth component to rapidly decrease outside this thin layer, at least one of components of the base field $\phi_a(r)$ must go to zero outside this layer. 

In the limit when the effective thickness of vorticity layer goes to zero the space integrals involving vorticity such as we have in Biot-Savart law and our net velocity, will be dominated by the delta term and stay finite.

\section{Finite Element Approximation}
Now, we assume that the function $H(\vec x)$ is a smooth function on a surface. Then the following numerical approach would work.

Let us cover the domain $D_C$ by a square grid step $1$ and assume that there are large number of these unit squares inside the loop. Let us approximate the loop by the loop drawn on this grid, passing through its cites. 

Eventually we shall tend the area of $D_C$ to infinity, in which case this quantization will become irrelevant.

Now let us approximate $H(\vec r)$ by its value at the center $\vec c_\Box$ inside each square $\Box$
\begin{equation}
    H(\vec r \in \Box) \approx h_\Box = H(\vec c_\Box)
\end{equation}

The resulting integral over the square is calculable:
\begin{equation}
    I_\alpha(\Box, \vec r) = \frac{1}{2 \pi}\int_{\Box} d^2 r' \dal' \frac{1}{|\vec r - \vec r'|}= \sum_{i=0}^3 (-1)^i A_\alpha\left(\vec V_i - \vec r\right)
\end{equation}
Here $\vec V_i, i =0,1,2,3$ are the vertices of $\Box$, counted anticlockwise starting with the left lowest corner $\vec V_0$ and 
\begin{equation}\label{A}
    A_\alpha(\vec r) = \frac{1}{2\pi} \arctanh \hat r_\alpha; \\ \hat r = \frac{\vec r}{|\vec r|};
\end{equation}

Thus we get an approximation
\begin{equation}
    F_\alpha[H,\vec r] \approx  \sum_{\Box \in D_C}h_\Box I_\alpha(\Box,\vec r)
\end{equation}

After that the target functional $Q[H]$ becomes an ordinary quadratic form of a vector $h_\Box, \Box \in D_C$.

The integral $\int_{D_C} d^2 r$ in \eqref{Target} converges (there is logarithmic singularity in $A_\alpha(\vec r)$ at $r_\alpha \ra \pm|\vec r|$, but it is integrable).
We have to compute symmetric matrix 
\begin{equation}
    \left<\Box_1 \left|M\right| \Box_2\right>  = \int_{D_C} d^2 r I_\alpha(\Box_1,\vec r) I_\alpha(\Box_2,\vec r) 
\end{equation}
and the linear term
\begin{equation}
     \int_{D_C} d^2 r R(\vec r) H(\vec r)  \approx \sum_{\Box} \bar R(\Box) h_{\Box}
\end{equation}
where $\Box_0$ is the square at the origin (the center of the domain).

These integrals for the matrix elements as well as the linear term are calculable with $5$ significant digits  using adaptive cubature library\cite{cubature}, based on recursive subdivision of the multidimensional cube\cite{Genz91}. We wrote parallel code which works fast enough for millions of squares on a supercomputer. 

For numerical stabilization we replaced the singular logarithm function in \eqref{A} by cutoff function at $\epsilon = 10^{-6}$
\begin{eqnarray}
    &&A_\alpha(\vec r) \approx \frac{\ln\left(1 + \hat r_\alpha,\epsilon\right) - \ln\left(1 - \hat r_\alpha, \epsilon\right)}{4\pi}; \\
    &&\ln(x,\epsilon) =  \ln\left(\max\left(|x|,\epsilon\right)\right)
\end{eqnarray}

We also added to our target the stabilizer:

\begin{eqnarray}
    &&Q[\vec h] = -\sum_{\Box}h_{\Box}\bar R(\Box) + \nonumber\\
    &&\oh\sum_{\Box_1,\Box_2} h_{\Box_1} \left<\Box_1 \left|M\right| \Box_2\right>  h_{\Box_2} +\nonumber\\
    &&\oh \lambda(M)\sum_{<\Box_1,\Box_2>} (h_{\Box_1} -h_{\Box_2})^2
\end{eqnarray}

Here  $\Box_0$ is the origin in our plane,  $<\Box_1,\Box_2>$ denote squares sharing a side and
\begin{eqnarray}
    \lambda(M) = \max |\delta M|
\end{eqnarray}
is maximal absolute error in computation of numerical integrals for matrix elements of $M$, in our case $\lambda \sim 10^{-6}$.

There are also two constraints (with $\vec 0$ representing the origin, which is a geometric center of the domain):
\begin{eqnarray}
    &&C_1: h_\Box  =0 ; \forall \Box \in C;\\
    &&C_2: \sum_\Box h_\Box I_\alpha(\Box,\vec 0) =0
\end{eqnarray}
Once the matrix $M$ is computed, the solution for the grid weights $h_{\Box}$ is given by  the minimum of quadratic form $Q$ with conditions $C_1,C_2$
\begin{equation}
    h_\Box = \argmin\left[Q\right]_{ C_1, C_2}
\end{equation}
As for the symmetric positive definite matrix inversion, there are fast parallel libraries\cite{PLASMA09}  available in $c^{++}$, so this looks achievable even for the grids with million squares.

We are planning to perform this computation for rectangles with various aspect ratios on a supercomputer and compare to available DNS data.

The circulation integral in terms of these coefficient $h_\Box$ reads
\begin{equation}\label{MinSol}
    \Gamma[C] = m \sum_{\Box} h_{\Box} \INT{0}{2 \pi} d \theta \int_{\Box} d^2 r\left(\frac{1}{\left|\vec r -L\vec f(\theta)\right|} - \frac{1}{\left|\vec r \right|}\right)
\end{equation}
Note that in virtue of our boundary condition $h_{\Box \in C} =0$ the singular terms with the squares at the boundary $C = \d D$ are excluded from the sum.

The remaining terms contain integrals over the angle $\theta$ of the double integrals $\int_{D_C} d^2 r$ of Coulomb kernel .

These integrals are calculable. The basic integral reads
\begin{subequations}
\begin{eqnarray}\label{CoulombInt}
   && B(x,y) \equiv \int_0^x  \int_0^y  \frac{d u d v}{\sqrt{u^2 + v^2}} =I(x,y) + I(y,x)\nonumber\\
   && I(x,y) = x \arcsinh\frac{y}{\sqrt{x^2 + \epsilon}}
\end{eqnarray}
\end{subequations}
The integral over the square $\Box(\vec P, \vec Q)$ with corners at $\vec P$ and $\vec Q $ is given by sum of four terms

\begin{eqnarray}
   &&G\left(\vec P, \vec Q\right) =\int_{\Box(\vec P, \vec Q)} \frac{d^2 r}{|\vec r|} =\nonumber\\
   &&B(Q.x,Q.y) - B(P.x,Q.y) - \nonumber\\
   &&B(Q.x,P.y) + B(P.x,P.y)
\end{eqnarray}

So, we represent the integral as (with $\vec C = \left(\oh a, \oh b\right)$ corresponding to the middle of the rectangle)
\begin{subequations}
\begin{eqnarray}
    &&\INT{0}{2 \pi} d \theta \int_{\Box(\vec P, \vec Q)} d^2 r \left(\frac{1}{\left|\vec r -L\vec f(\theta)\right|} - \frac{1}{\left|\vec r - \vec C \right|}\right) =\nonumber\\
    &&J_x + J_y - 2 \pi  G\left(\vec P-\vec C, \vec Q-\vec C\right);\nonumber\\
    && J_x = \int_{-a/b}^{a/b} d t \frac{G\left(\vec P(t,0),\vec Q\right) + G\left(\vec P(t,b), \vec Q(-b/a,b)\right)}{1 + t^2}\nonumber\\
    && J_y = \int_{-b/a}^{b/a}d t \frac{G\left(\vec P, \vec Q(t,0)\right) + G\left(\vec P(-a/b,a), \vec Q(t,a)\right)}{1 + t^2} \nonumber\\
    && \vec P(t,c) = \vec P - \left(\frac{a + b t}{2},c\right)\\
    && \vec Q(t,c) = \vec Q - \left(\frac{b + a t}{2},c\right)
\end{eqnarray}
\end{subequations}

These $\vec P(t), \vec Q(t)$ are equations of the sides of our polygon.
Also note that in the limit of large size of the domain, when the number $N$ of grid squares goes to infinity, the coefficients $h_\Box$ decrease as $1/N$. 

In this limit, our sum over squares becomes the Riemann sum for an integral \eqref{Falpha}.

The reason for exactly computing the integrals over elementary squares with constant $H(\vec r)$ inside each square was the Coulomb singularity. Resulting functions $A_\alpha(\vec r), B(x,y)$ has only a logarithmic singularities, rather than the pole in Coulomb potential. So, the integrals involving these functions can be computed with high accuracy using cubature package\cite{cubature} using
regularization of logarithms with $\epsilon$ terms.

By exactly computing singular integrals we accelerated the convergence to a local limit $N \ra \infty$. With Riemann sums for Coulomb kernel the errors would be $O\left(1/\sqrt{N}\right)$, but with replacing $H(\vec r)$ by its values at the center the relative errors are related to second derivatives which is $O\left(1/N\right)$. So, with accessible $N \sim 10^6$ at modern supercomputers we expect to get $5$ significant digits, which is beyond the statistical and systematic errors of the DNS at achievable lattices  $24K^3$.

The hardest part of this computation is numerical integration needed for the kernel $\left<\Box_1 \left|M\right| \Box_2\right> $ for all the squares $\Box_1, \Box_2$. It has $O\left(N^3\right)$ complexity where $N$ is the number of squares inside $D_C$. Still, with $N\sim 100$ this (parallel) computation using adaptive cubature library\cite{cubature} takes less than a minute on my server with $24$ cores.

\bibliography{bibliography}

\end{document}